\documentclass{aa}

\usepackage[varg]{txfonts}

\usepackage{amsmath}

\DeclareRobustCommand{\VAN}[3]{#2}
\let\VANthebibliography\thebibliography
\def\thebibliography{\DeclareRobustCommand{\VAN}[3]{##3}\VANthebibliography}

\usepackage{graphicx}	
\usepackage{amsmath}	

\begin{document}

\title{Supernova shocks cannot explain the inflated state of hypervelocity runaways from white dwarf binaries}

\author{Aakash Bhat\inst{1,2,3}
        \and
        Evan B. Bauer\inst{4,3}
        \and
        R\"udiger Pakmor\inst{3}
        \and
        Ken J. Shen\inst{5}
        \and
        Ilaria Caiazzo\inst{6,7}
        \and
        Abinaya Swaruba Rajamuthukumar\inst{3}
        \and
        Kareem El-Badry\inst{7}
        \and
        Wolfgang E. Kerzendorf\inst{8,9}
        }
\institute{Institut für Physik und Astronomie, Universität Potsdam, Haus 28,            Karl-Liebknecht-Str. 24/25, 14476 Potsdam-Golm, Germany\\
              \email{aakashbhat7@gmail.com}
        \and
            Dr Karl Remeis-Observatory \& ECAP, Friedrich-Alexander University Erlangen-Nürnberg, Sternwartstr. 7, 96049 Bamberg, Germany
        \and
            Max Planck Institut für Astrophysik, Karl-Schwarzschild-Straße 1, 85748 Garching bei München, Germany
        \and
            Center for Astrophysics | Harvard \& Smithsonian, 60 Garden Street, Cambridge, MA 02138, USA
        \and
            Department of Astronomy and Theoretical Astrophysics Center, University of California, Berkeley, CA 94720-3411, USA
        \and
            Institute of Science and Technology Austria (IST Austria), Am Campus 1, Klosterneuburg, Austria
        \and
            Division of Physics, Mathematics and Astronomy, California Institute of Technology, 1200 E. California Blvd., Pasadena, CA 91125, USA
        \and
            Department of Physics and Astronomy, Michigan State University, East Lansing, MI 48824, USA
        \and
            Department of Computational Mathematics, Science, and Engineering, Michigan State University, East Lansing, MI 48824, USA
            }

\abstract{

Recent observations have found a growing number of hypervelocity stars with speeds of $\approx 1500-2500\,$km\,s$^{-1}$ which could have only been produced through thermonuclear supernovae in white dwarf binaries. Most of the observed hypervelocity runaways in this class display a surprising inflated structure: their current radii are roughly an order of magnitude greater than they would have been as white dwarfs filling their Roche lobe. While many simulations exist studying the dynamical phase leading to supernova detonation in these systems, no detailed calculations of the long-term structure of the runaways have yet been performed. We use an existing \textsc{Arepo} hydrodynamical simulation of a supernova in a white dwarf binary as a starting point for the evolution of these stars with the 1 dimensional stellar evolution code MESA. We show that the supernova shock is not enough to inflate the white dwarf over timescales longer than a few thousand years, significantly shorter than the $10^{5-6}$ year lifetimes inferred for observed hypervelocity runaways.
Despite experiencing a shock from a supernova less than $\approx 0.02\,R_\odot$ away, our models do not experience significant interior heating, and all contract back to radii around $0.01\,R_\odot$ within about $10^4$\,years. Explaining the observed inflated states requires either an additional source of significant heating or some other physics that is not yet accounted for in the subsequent evolution.
}


\maketitle



\section{Introduction}

Thermonuclear supernovae in white dwarf binaries can produce a hypervelocity runaway companion, and we are now discovering some of the fastest stars in our Galaxy that appear to have originated from this scenario \citep{2018ApJ...865...15S,2023arXiv230603914E}. While the growing number of observed candidates from such a scenario seems to confirm the importance of double white dwarfs as thermonuclear supernova progenitors, the stellar structure of these runaways is still poorly understood. This leaves an incomplete picture of the connection between these runaways in our Galaxy and the population of Type Ia supernovae observed in our Universe. In the "dynamically-driven double degenerate double detonation" (D$^6$) scenario, a white dwarf transfers mass to a more massive white dwarf through dynamically unstable Roche-lobe overflow \citep{Guillochon,Dan2011,Dan2012,2013ApJ...770L...8P,Shen2018a}. This unstable mass transfer, mainly comprised of heavily compressed $^4$He, leads to a thermonuclear runaway through dynamical instabilities in the interaction between the accretion stream and the shell of the accretor. This can trigger a detonation. The shell detonation leads to a shock that traverses inside the white dwarf and converges in the carbon-oxygen core. At this stage, it is possible to ignite carbon and detonate the white dwarf  \citep{2007A&A...476.1133F, 2010ApJ...719.1067K,2013ApJ...770L...8P,2013ApJ...774..137M,2021ApJ...919..126B,2022MNRAS.517.5260P}. Once this happens the accreting white dwarf effectively disappears and the donor star is ejected at its pre-supernova orbital speed and becomes a runaway star. Due to the compact nature of the stars, the secondary is very close to the primary before filling its Roche lobe. Therefore the runaways can reach velocities of the order $1000-2500$ km\,s$^{-1}$. For runaways observed with velocities above $\approx 1500$ km\,s$^{-1}$ that are not coming from the Galactic centre (which could indicate a \citealt{Hills1988} mechanism origin), the D$^6$ supernova scenario is the only way to explain their extreme velocities without relying on objects like intermediate mass black holes. 

 All the observed D$^6$ runaway candidates are peculiar. They have galactocentric velocities greater than $1000 \,\mathrm{km\,s}^{-1}$, with the fastest ones being $\approx2500 \,\mathrm{km\,s}^{-1}$. As they are unbound to the Galaxy and still observable, they can not be more than $10-100$~Myr removed from the supernova event. Moreover, their observed flight times to the Galactic disk -- and in at least one case, to SN remnants -- suggest lifetimes of at most a few Myr. Due to the mass-radius relation of white dwarfs, these velocities directly imply that the donor white dwarfs must have been massive ($>1\,$ M$_\odot$) for the fastest runaways and ($>0.5\,$ M$_\odot$) for the slightly slower ones \citep{2021ApJ...923L..34B}. While the difference in composition of sub-Chandrasekhar mass white dwarfs can affect the mass-radius relation by a few percent and lead to slightly different orbital velocities, these differences will be less than the uncertainties in observed orbital velocities due to uncertain distances. We also note the caveat that the preceding mass estimates assume a Roche-filling donor at the time of explosion. It may also be possible that the accretor detonates later when the donor is significantly beyond its Roche limit, so that the plunging donor star might reach higher velocities than when it fills it Roche-lobe. In this case the mass of the runaway could be lower, down to $0.4$ M$_\odot$ \citep{Dan2012}. In any case, almost all of the observed white dwarfs are puffed up, with  most having radii on the order of 0.1 R$_\odot$ instead of $<0.01$ R$_\odot$ as expected from their velocities. One possible explanation that has been suggested for this behaviour is post-shock induced heating due to the supernova, and by supernova ejecta material deposited on the surface. So far, some stellar evolution models have addressed the impact of supernova shocks on ejected subdwarfs in single-degenerate supernovae \citep{2019ApJ...887...68B}, double-degenerate supernovae with a helium white dwarf donor \citep{sunny2024}, and Type Iax remnant stars \citep{2019ApJ...872...29Z}, to explain observations of those runaways \citep{2005A&A...444L..61H,2015Sci...347.1126G,Vennes2017,Raddi2018,2019MNRAS.489.1489R}.

While these efforts have provided insights into the evolution of subdwarfs and partially burned white dwarfs, a detailed investigation into the evolution of the faster D$^6$ runaways is still missing. As we will show in this work, and as estimated analytically in \cite{2019ApJ...887...68B}, it is much more difficult for the supernova shock to strongly perturb the thermal state of a D$^6$ WD donor than in the previously studied cases of subdwarf donors or Iax remnants. In this paper, we improve upon previous work by carrying out the first full-scale stellar evolution calculation of such surviving white dwarfs. We utilise the result of an existing \textsc{Arepo} 3-dimensional hydrodynamical simulation of binary white dwarfs which undergo a supernova explosion and end with a hypervelocity runaway donor remnant \citep{Arepo, Pakmor2016, 2020ApJS..248...32W, 2022MNRAS.517.5260P}. This simulation taken from \citet{2022MNRAS.517.5260P} followed the evolution until shortly after double detonation of the primary white dwarf. In total the runaway was evolved for $150$ s after explosion when the ejecta are in homologous expansion and the surviving secondary white dwarf has settled into hydro-static equilibrium. We take this surviving donor as input into the 1-dimensional open-source stellar evolution software Modules for Experiments in Stellar Astrophysics (MESA, \citealt{Paxton2011,Paxton2013,Paxton2015,Paxton2018,Paxton2019,Jermyn2023}). 

The \textsc{Arepo} simulation provides profiles of the composition and thermal state of pre- and post-supernova configurations for the donor, which becomes a hypervelocity runaway. We use the entropy profiles to calculate the heating associated with the supernova shock. We use the result to create an approximate model of the heating for different masses of white dwarfs (0.5 to 1.1 M$_\odot$) to explore how the subsequent evolution may depend on runaway mass. We then model the subsequent stellar evolution through the following $100\, \mathrm{Myr}$, after which time these stars would be in the outer halo of the Galaxy and not observable. The mapping of only one available \textsc{Arepo} simulation is not exact when extrapolated to white dwarf binaries of other masses. The heating of a donor white dwarf likely has some dependence on mass, since a more massive white dwarf will be closer to the supernova, but at the same time is more difficult to perturb due to its higher pressure and internal energy profile. As a first exploration based on the one available \textsc{Arepo} model, our work scales the heating from the 0.7~$M_\odot$ simulation to higher masses, representing a maximally aggressive heating to place an upper limit on how long the white dwarfs can stay puffed up. In Section~2, we discuss the \textsc{Arepo} simulations and develop an approximate approach to generalise the heating we see in the one \textsc{Arepo} model to white dwarf donors of different masses. In Section 3, we present the evolution of all our models. Finally, Section 4 compares these models to recent observation of D$^6$ runaways. We summarise our results in Section 5.

\section{Simulations and Relaxation}

Our goal in this work is to model the long-term stellar evolution of the runaway donor in order to compare it with observed hypervelocity runaways. These runaways travel at speeds on the order of kpc/Myr, so we expect that they would leave the Galaxy and become unobservable within about 10~Myr. We therefore want to compare evolutionary tracks with ages reaching a few Myr after explosion for typical observable runaways, though some could be younger. After the initial interaction with the SN ejecta and dynamical evolution over timescales of a few hundred seconds modelled in \textsc{Arepo}, the donor star should quickly settle \citep{Shen2012,Schwab2012} into an approximately spherical and hydrostatic structure appropriate for modelling with a 1D stellar evolution code such as MESA for longer timescales. We neglect rotation in this work because even a tidally locked donor star will be significantly below critical rotation, and the energy in rotational shear in the \textsc{Arepo} model is negligible compared to other energy scales. In this section, we describe the initial dynamical evolution modelled in \textsc{Arepo} up to a point suitable to handing off to MESA for longer term evolution, along with the procedure for mapping from the 3D Arepo model into the eventual 1D hydrostatic structure for MESA.

\subsection{3D hydrodynamical explosion model}
We start from an existing explosion simulation in \textsc{Arepo} of a double white dwarf system \citep{2022MNRAS.517.5260P}. This simulation starts with the late stages of inspiral of a binary system of two carbon-oxygen white dwarfs with masses of $1.05\,\mathrm{M_\odot}$ and $0.7\,\mathrm{M_\odot}$, respectively. The donor model starts with a 50/50 C/O core composition and a pure He envelope in the outer 0.03\,$M_\odot$. This helium mass is significantly higher than the mass required to support a double detonation \citep{Shen2024}, and almost an order of magnitude more than the helium shell masses of realistic WD models (e.g., \citealt{2006MNRAS.371..263L}). This is motivated by the need to have enough He mass transfer such that the accretor is sure to double-detonate. Accretion of helium-rich material from the less massive donor to the more massive accretor dynamically ignites a helium detonation in the helium shell of the accretor. The helium detonation propagates around the accretor and via compression ignites a detonation in the carbon-oxygen core of the accretor, similar to the classic double detonation scenario for Type Ia supernovae. 

The accretor explodes as a Type Ia supernova. It disappears on a timescale of a few seconds, and hits the donor white dwarf with a strong shockwave on the way, that travels with a velocity of $\sim 10{,}000$ km\,s$^{-1}$. During this process, the donor white dwarf also burns its helium shell and loses its outer layers. If it survives the impact and does not also explode itself \citep{2019ApJ...885..103T, 2022MNRAS.517.5260P, boos2024, Shen2024}, new material (in particular $^{56}$Ni) is deposited from the low-velocity tail of the supernova ejecta, resulting in a final mass of the donor of $0.67\,$M$_{\odot}$. While the donor experiences a shock from the interaction with supernova ejecta, this shock is only strong enough to deposit significant entropy near the surface. The \textsc{Arepo} model also experiences some additional viscous heating during post-shock ringdown, but again, this is negligible in the core.

Due to its almost Lagrangian nature, \textsc{Arepo} can follow both the expanding supernova ejecta and the resolved surviving donor white dwarf for more than $100\,\mathrm{s}$ after the explosion. Here we use the surviving donor of this existing simulation as a starting point to study its subsequent evolution. We employ the composition and entropy profiles of the donor at the end of the \textsc{Arepo} simulation and insert them into MESA. Our donor is on the less massive end of the masses predicted from roche-lobe filling donors based on observed runaway velocities. In this work we will also use the heating profile from this model to estimate the potential impact of supernova interaction for more massive WD donors as well. At this stage, creating a new simulation with more massive white dwarfs was not within the scope of this work. In reality, we expect that more massive WDs would be even harder to perturb with supernova shocks due to higher internal pressures \citep{2019ApJ...887...68B}, and therefore our work represents an upper limit on supernova shock heating for the more massive models.

\begin{figure}
\hspace*{-0.3cm}
\centering
\includegraphics[width=0.5\textwidth]{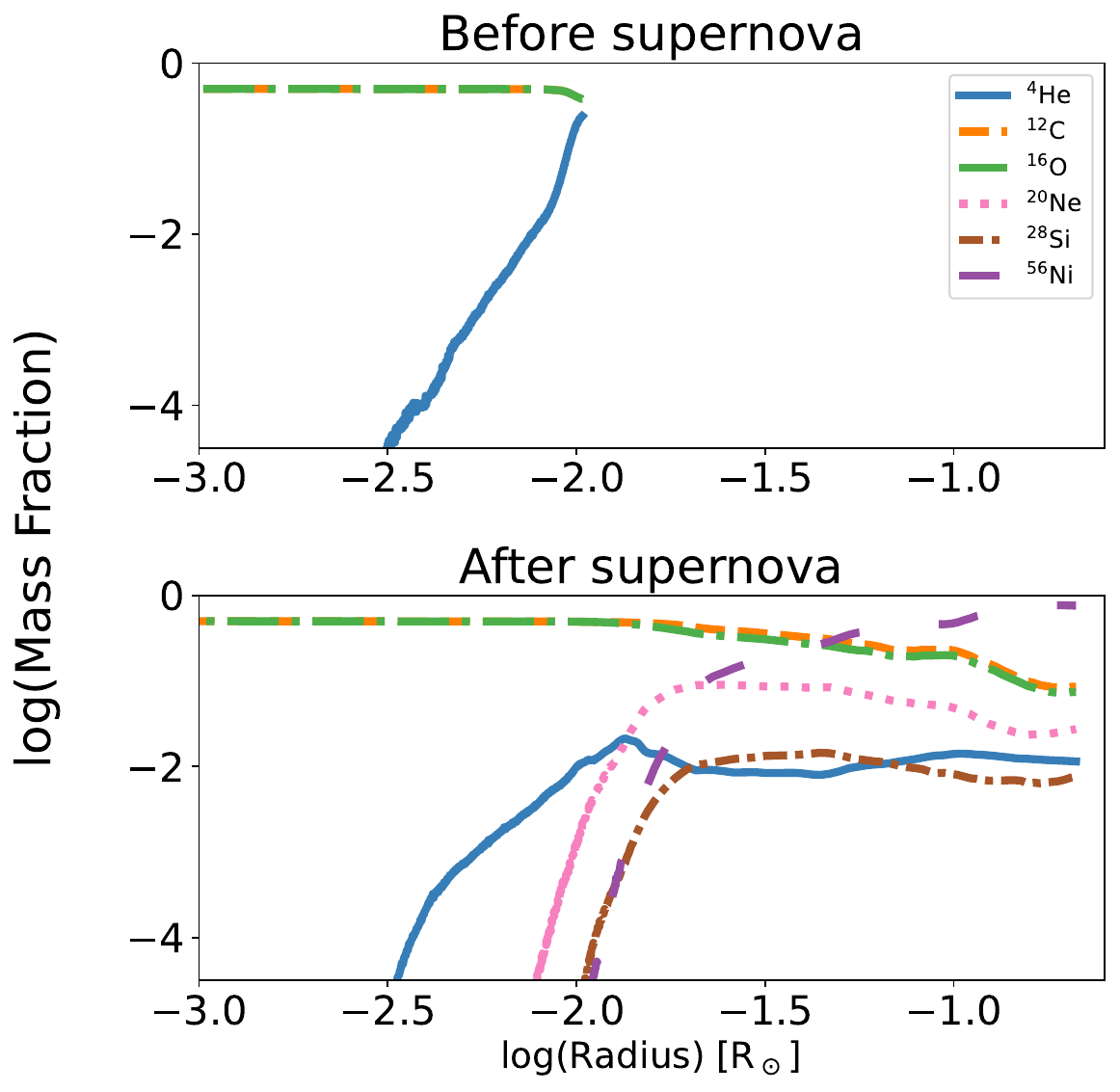} 
\caption{Main elements and their composition profiles as a function of radius for the donor right before the supernova \textbf{[top]} and the runaway 150~s after the supernova \textbf{[bottom]}  for mass coordinates which remain bound after supernova explosion ($<0.661\, M_\odot$). In particular, $^{56}$Ni dominates the mass fraction at the surface of the runaway. The surface helium content of the donor is less than 0.03 M$_\odot$ as most of it has been accreted by the primary.  }
\label{fig:compo}
\end{figure}

The final \textsc{Arepo} profile of the remnant contains composition information and spherically averaged 1-D temperature-density pairs for different shells of the star. This profile differs from the initial profile of the donor due to the presence of heavy elements generated during the supernova explosion. The final composition profiles of the main elements with mass fractions $\max(X_n) > 10^{-4}$ within the donor and the remnant are shown as a comparison in Figure \ref{fig:compo}. 

\begin{figure}
\centering
\includegraphics[width=0.45\textwidth]{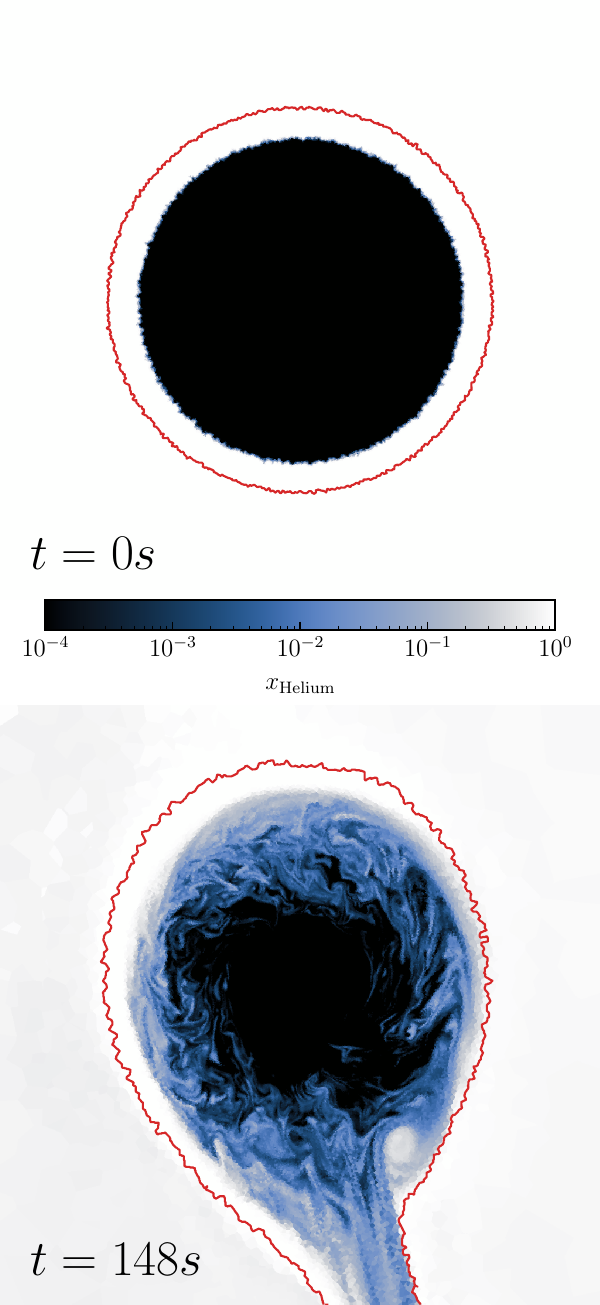}
\caption{\textbf{[top panel]} Helium mass fraction at the beginning of the simulation showing no mixing. \textbf{[bottom panel]} Mixing of helium in the donor just before the explosion of the accretor. The red line shows a contour of $\rho=10^4\,\mathrm{g\,cm^{-3}}$. During the accretion and inspiral the helium from the surface seeps into the deeper layers of the donor.}
\label{fig:hemix}
\end{figure}

Notably, there is a tail of helium from the envelope that is mixed deeply into the core of the surviving white dwarf. Figure~\ref{fig:hemix} shows slices through the midplane of the donor white dwarf at the beginning of the binary simulation and just before the accreting white dwarf explodes in a supernova. The mixing of helium into its core is driven by shear on the surface. However, the refinement and derefinement in the \textsc{Arepo} moving mesh in this simulation during inspiral approaching mass transfer significantly enhances this shear mixing in an nonphysical manner. We checked this by changing the refinement of the donor surface separately (without re-running the entire simulation) and found the mixing to be negligible. Experimenting with the addition of resolution, we found the mixing to essentially completely disappear. While we can't entirely exclude the possibility that some shear mixing could occur over much longer timescales than the dynamical \textsc{Arepo} simulation, these experiments confirm that the mixing seen in Figure~\ref{fig:hemix} is a numerical artefact. Since we believe in particular the deep mixing is mostly artificial, we will evolve models of the surviving white dwarf where we remove this helium tail from the core by replacing any $^4$He with $^{12}$C in the inner 99\% of the star by mass coordinate. The evolution of the original \textsc{Arepo} composition is provided in Appendix \ref{B.Arepo}.

\subsection{Mapping the hydrodynamical result into 1D stellar evolution}

To model the long-term evolution of the remnant runaway, we evolve it further in MESA \citep{Paxton2011, Paxton2013, Paxton2015, Paxton2018, Paxton2019, Jermyn2023}, an open-source 1-dimensional stellar evolution code. The components of the MESA EOS blend relevant for this work are FreeEOS \citep{Irwin2004} near the surface of models, and HELM \citep{Timmes2000} and Skye \citep{Jermyn2021} for white dwarf cores.
MESA uses tabulated radiative opacities primarily from OPAL \citep{Iglesias1993,
Iglesias1996}, with low-temperature data from \citet{Ferguson2005}
and the high-temperature, Compton-scattering dominated regime by
\citet{Poutanen2017}. The electron conduction opacities are from
\citet{Cassisi2007}. Nuclear reactions are from JINA REACLIB \citep{Cyburt2010}, NACRE \citep{Angulo1999} and additional tabulated weak reaction rates \citet{Fuller1985, Oda1994,Langanke2000}. Screening of reaction rates is included via the prescription of \citet{Chugunov2007}, while thermal neutrino loss rates are from \citet{Itoh1996}.

In order to map the change in thermal state of the white dwarf donor into MESA, we map the change in {\em entropy} that the donor star experiences as a result of interaction with the supernova ejecta. This is important, because when a supernova occurs in a close binary system, the ejecta interact with the donor in several ways. They cause a shock to propagate through the donor and directly deposit thermal energy in the interior. The ejecta also strip away material from the surface of the star. Moreover, they shock the helium shell of the donor white dwarf, which detonates and burns \citep{2022MNRAS.517.5260P}. Here we assume that the core of the donor does not explode as well \citep{2022MNRAS.517.5260P,Shen2024}. Essentially all of the ashes of the helium burning on the surface of the donor will be ejected. Interior fluid elements experience shock heating that can increase their entropy, followed by adiabatic decompression to a lower density configuration as the donor white dwarf settles into a less compact configuration at lower mass. Therefore, the specific internal energy $e$ of a white dwarf layer experiences several different terms contributing to its evolution, and in general may end up {\em lower} than its pre-shock value due to stripping and eventual decompression. In contrast, the specific entropy will still be higher and encode the irreversible non-adiabatic heating that occurred during its interaction with the supernova ejecta. We therefore calculate entropy profiles of the donor white dwarf in \textsc{Arepo} just before the supernova explosion and after the donor has settled into a well-defined, essentially spherical, self-gravitating runaway star at the end of the hydrodynamical simulation. The latter state is 150 s after the supernova explosion. For comparison, the dynamical timescale, $t_{\mathrm{dyn}}=\sqrt{R^3/(GM)}$, for the bulk of the interior is less than 10\,s. The outer few percent of the mass of the runaway that was accreted from SN ejecta is more inflated and has a longer dynamical timescale on the order of 100\,s, but as we will show, most of this material will quickly become unbound. 

Since MESA is a 1-D evolutionary code, we averaged the 3-D profiles from \textsc{Arepo} over spherical shells defined with respect to the centre of mass of the surviving bound donor. In the case of directly modelling the evolution of the runaway remnant from the \textsc{Arepo} model, we can use the MESA relaxation methods to initialise a WD model directly to the entropy and composition profile specified by the final state of the \textsc{Arepo} model. As we only have one simulation, we also develop an approximate generalised model to account for shock heating in MESA WD models of other masses. The benefit of our method is that we can use the same heating method for multiple white dwarfs of different masses.

\begin{figure}
\centering
\includegraphics[width=0.5\textwidth]{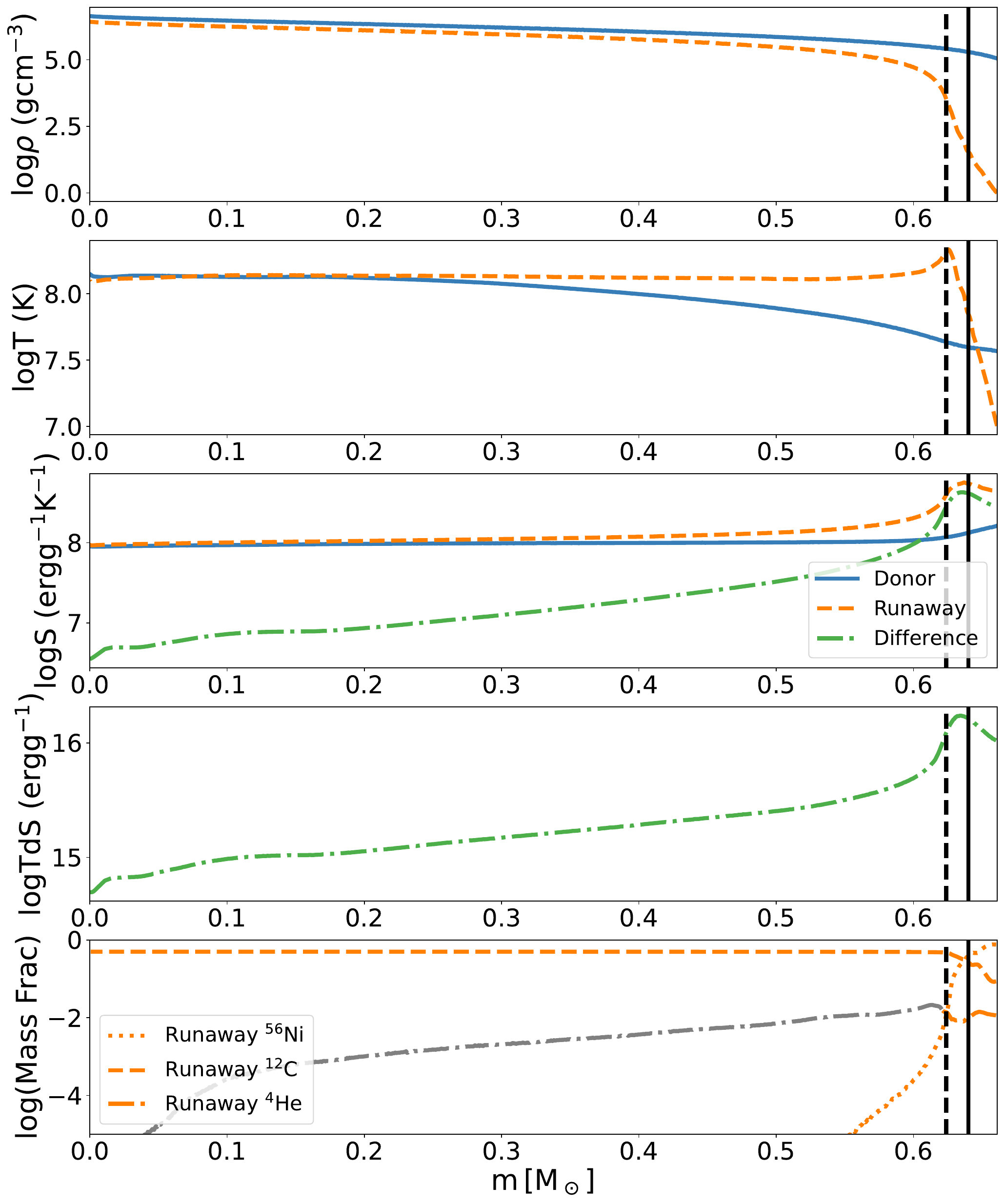} 
\caption{from the top, \textbf{[first panel]} density profiles of the donor (blue) and runaway star (orange). \textbf{[second panel]} Temperature profiles of the donor and runaway. \textbf{[third panel]} Entropy profiles of the donor and the runaway. \textbf{[fourth panel]}  The non-adiabatic heating done after the ejecta hits. \textbf{[fifth panel]} The mass fractions of $^{12}$C (similar to $^{16}$O), $^4$He, and $^{56}$Ni. The mass coordinates, given in solar masses, are shown until the donor mass of $0.661\, M_\odot$, as outside of this mass range entropy difference is not defined. The dashed black line represents the mass coordinate where $^{12}$C and $^{16}$O mass fractions become less than 90\% of the total mass fraction. The solid black line shows where the surface cut as defined in section \ref{relaxheat} is done. The grey portion for helium marks what we remove in our standard models. }
\label{fig:entropy_comp}
\end{figure}

To calculate the input heat required to achieve the desired entropy change, we first calculate the entropies of the \textsc{Arepo} white dwarfs (we call the secondary white dwarf a donor before the supernova and a runaway star after the supernova). We specifically start from the \textsc{Arepo} profiles of temperature, density, and composition. From them we compute entropy profiles using PyMesa \citep{pymesa} to evaluate the EOS entropy as a function of density, temperature, and composition. We set the equation of state as HELM \citep{2000ApJS..126..501T} everywhere within the star, consistent with the EOS used in the \textsc{Arepo} simulations. The middle panel in Figure~\ref{fig:entropy_comp} shows the entropy profiles of the donor white dwarf before and after the supernova, and the difference in entropy which can be provided in MESA as a heating term. 

For a given amount of input heat $\delta Q$, the first law of thermodynamics states that the entropy and energy will change according to $\delta Q = TdS = dE + PdV$.
Our goal is to achieve the same non-adiabatic heating profile $T\Delta s$ represented by the hydrodynamics of the \textsc{Arepo} model. For the case when temperature profiles of our MESA models and the \textsc{Arepo} model are roughly equal (as will be the case for the modelling in this work), we can apply the following heating term to achieve the desired entropy change:
\begin{equation}
    \epsilon_{\rm{heat}}= \frac{ T\Delta s_{\rm Arepo}}{\Delta t}~.
    \label{eq:heat3}
\end{equation}

According to the first law of thermodynamics, with specific heating rate $\epsilon_{\rm heat} = \delta q/\delta t$ (where $q$ is heat $Q$ per unit mass), the entropy change $\Delta s$ achieved for a MESA model over a time period of $\Delta t$ is then

\begin{equation}
    \Delta s = \int_0^{\Delta t}\frac{\epsilon_{\rm heat}\delta t}{T}~,
    \label{eq:heat1}
\end{equation}
and using the heating term defined in equation~\eqref{eq:heat3} in equation~\eqref{eq:heat1} then verifies that the total entropy change will be $\Delta s = \Delta s_{\rm Arepo}$, as desired. We verify that this heating approach produces the desired final entropy profile in our MESA models in the next section. 

The heating method is similar to one applied in \citet{2019ApJ...887...68B} for subdwarfs, where an entropy profile is reached by heating up the MESA model.  Unlike the Athena++ simulations used in \citet{2019ApJ...887...68B}, the \textsc{Arepo} simulations of \cite{2022MNRAS.517.5260P} track the composition evolution of the donor and include some amount of accreted supernova ejecta as part of the final bound remnant runaway. Therefore, we use the MESA composition relaxation to relax our models to the \textsc{Arepo} composition profiles first, and then heat them up using the  {\tt other\_energy} hook in MESA. For all evolution and relaxation scenarios, we use the {\tt approx21\_plus\_co56} nuclear network. This includes the isotopes $^{1}$H, $^{3,4}$He, $^{12}$C, $^{14}$N, $^{16}$O, $^{20}$Ne, $^{24}$Mg, $^{28}$Si, $^{32}$S, $^{36}$Ar, $^{40}$Ca, $^{44}$Ti, $^{48,56}$Cr, $^{52,54,56}$Fe, $^{56}$Co, and $^{56}$Ni. For our purpose the network therefore includes the most important reactions of nickel decay and $\alpha$ capture on carbon. 

\subsubsection{Direct Relaxation vs Heating}
\label{relaxheat}

We first test the heating approach by comparing our heated model to the relaxed one for the case where we can apply both procedures. This comparison requires three steps. First, we start by creating white dwarfs following the test suite template {\tt make\_co\_wd} in MESA. We then relax them to the donor profile and the remnant profile of the \textsc{Arepo} simulation. In the second step, we heat the donor white dwarf and compare the state after heating to the runaway white dwarf. In the final step, we compare the evolutionary tracks over $\mathrm{Myr}$ of evolution (discussed in Appendix~\ref{A.Comparison}).

\begin{figure}
\centering
\includegraphics[width=0.45\textwidth]{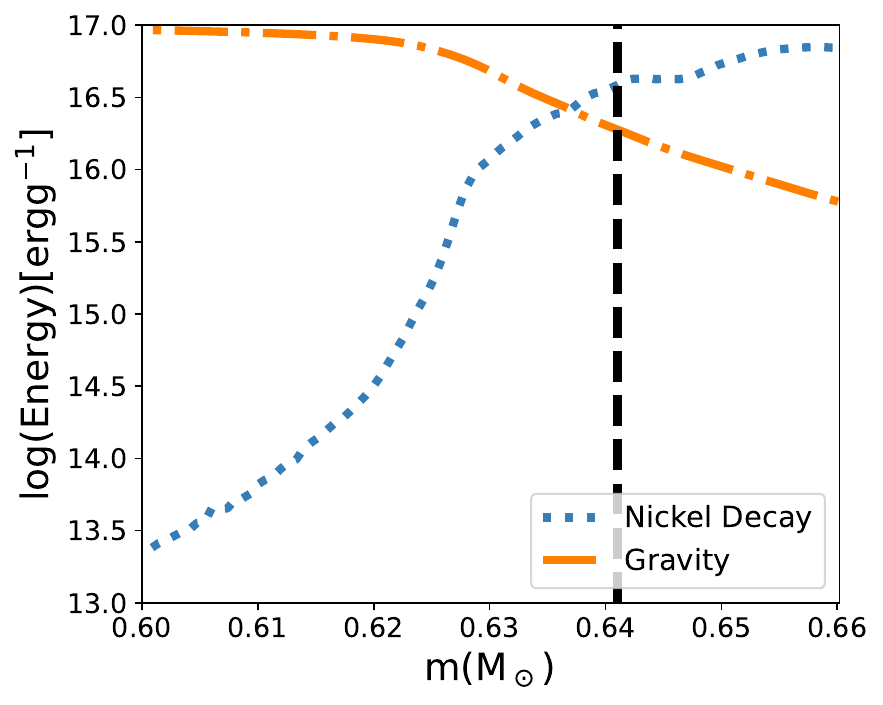} 
\caption{Comparison of $^{56}$Ni decay and the gravitational binding energy. The black dashed line marks our choice for the surface cut, where decay energy is twice the amount of binding energy. }
\label{fig:decay}
\end{figure}

We found it difficult at this stage to evolve models with significant amounts of $^{56}$Ni on the surface in MESA due to limitations in the EOS for regions with large fractions of high atomic number elements at low temperature. However, examining the overall energy scales at the surface reveals that the energy input from decay of $^{56}$Ni will quickly unbind much of the surface $^{56}$Ni on short timescales as soon as the heating from this decay can begin \citep{Shen2017}. We therefore estimated an outer layer from which we can confidently remove this $^{56}$Ni-dominated material as unbound to make the MESA model easier to evolve, while leaving enough of the outer accreted $^{56}$Ni, so that we will still see a residual outflow from nuclear decay energy at early times in our models. For this purpose, we defined an outer boundary for the bound material as the mass coordinate of the fluid element where the energy release due to $^{56}$Ni decay is twice the binding energy due to gravity. Gravitational binding energy (per unit mass) is defined as $Gm/r$, while we estimated Ni decay energy as $(5.38\,{\rm MeV})X_{\rm Ni}/(56m_{\rm p})$ from the total energy of the decay chain $^{56}$Ni$\rightarrow ^{56}$Co$\rightarrow ^{56}$Fe.  This yields a final mass of $0.64\,$M$_{\odot}$, where $0.021\,$ M$_{\odot}$ is the mass lost due to our surface cut. Out of this, we go from having 0.015 M$_{\odot}$ to $\approx 0.003$ M$_{\odot}$ of nickel. The binding energy and the decay energy is shown in Figure \ref{fig:decay} along with the surface cut allowing us to relax the WD successfully. The remaining $^{56}$Ni will still deposit enough radioactive decay energy to drive a residual outflow, as shown in the next section.

\begin{figure}
\centering
\includegraphics[width=0.47\textwidth]{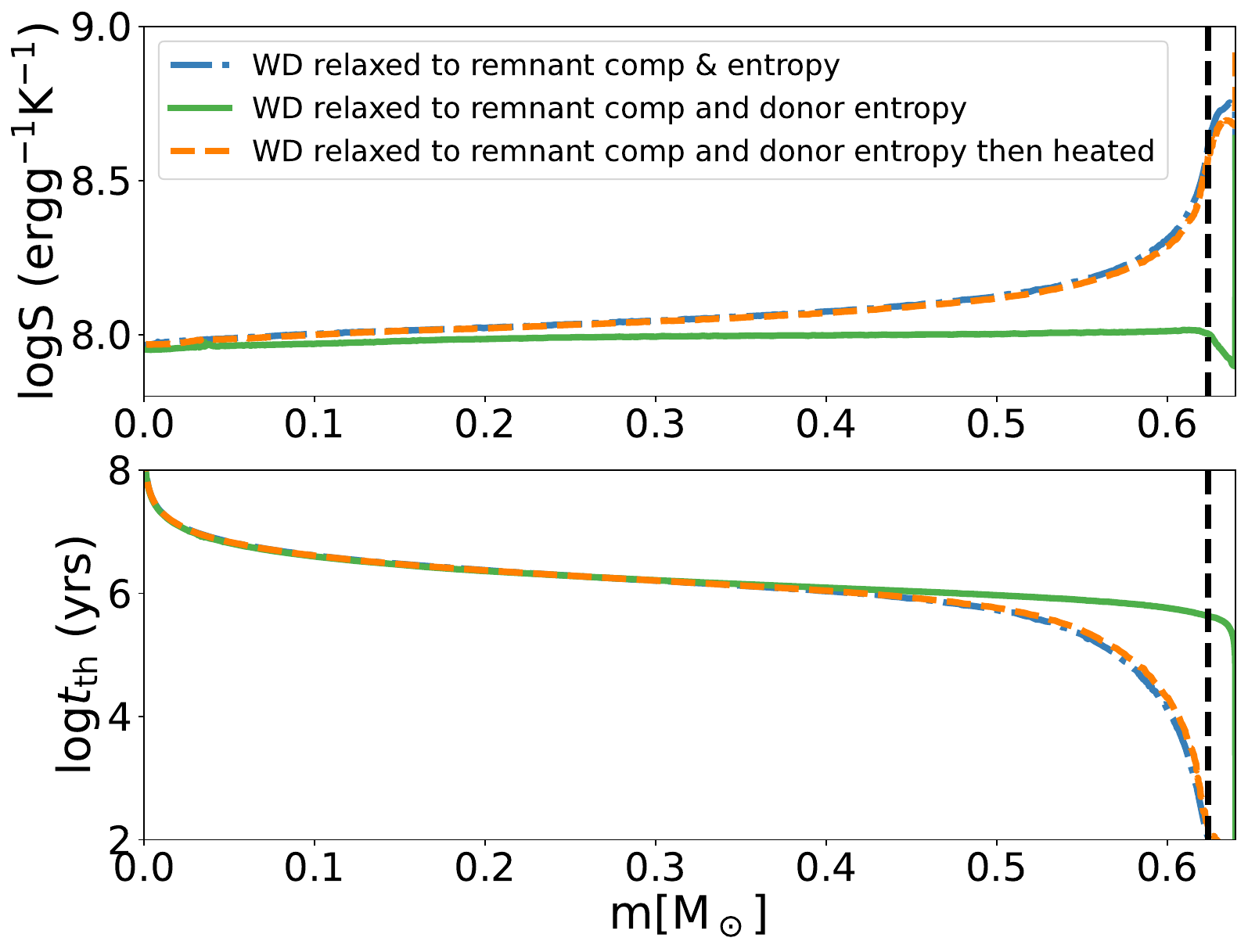} 
\caption{Comparison of heating with relaxation. There are three models here. The blue curve represents the model which has been relaxed to \textsc{Arepo} donor entropy and runaway composition. The orange curve represents the model which has been relaxed to runaway entropy and runaway composition. The green curve represents the blue model which has been heated. The green and orange curves overlapping shows that heating the white dwarf produces the same result as the relaxation. The dashed line marks the layers where the atmosphere is less than 90\% C-O dominated, and where an entropy difference is not well defined.}
\label{fig:heat_comp}
\end{figure}

For the second step, we provide the input heat as calculated in equation \eqref{eq:heat3} over an interval of $0.01$ seconds to the relaxed donor WD. During the heating we turned off nuclear burning and mixing. 
We stopped the evolution of the heated model after the desired amount of heating had been injected to compare to the results from the direct relaxation method. Figure \ref{fig:heat_comp} shows the entropy and local thermal diffusion time profiles for the relaxed and heated model. The local thermal diffusion time is given as
\begin{equation}
    t_{\rm{th}}=\frac{H^2}{D_{\rm{th}}}~,
    \label{eq:td}
\end{equation}
where $H=P/\rho g$ is the local pressure scale height and $D_{\rm{th}} = 4 a c  T^3/3 \kappa\rho^2 c_P$ is the coefficient of thermal diffusion.

The dashed black line marks the region where the atmosphere is nickel dominated (defined as the mass fraction of carbon + oxygen less than 90\%). Outside this region, the fluid elements do not represent entropy changes from heating, but rather are entirely different fluid elements from regions that were stripped and then replaced by the ejecta. The rest of the star shows almost perfect agreement in the final states. The evolutionary tracks for these models shown in Appendix~\ref{A.Comparison}  also show nearly perfect agreement except for small differences in the first few years of evolution governed by the transient evolution of the very outer layers as they are lost in a super-Eddington wind. The subsequent stellar evolution of these models agrees very well regardless of whether the entropy change is applied through direct relaxation or our heating procedure.

\subsubsection{Heating: other white dwarf masses}

\begin{figure}
\centering
\includegraphics[width=0.455\textwidth]{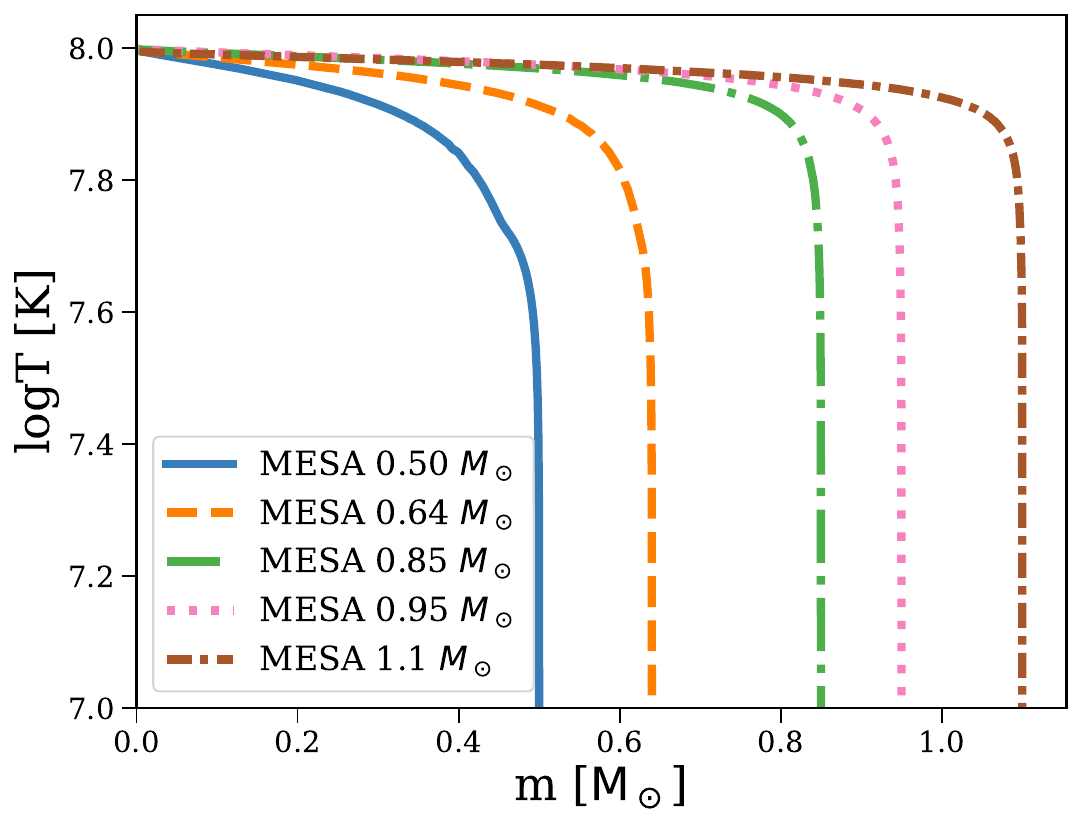} 

\caption{Temperature profiles of all WD models used in this work.}
\label{fig:wd_temp}
\end{figure}

Having established that the heating procedure produces a well-defined entropy change, we now explore a wider range of masses. We do this by creating a range of WD donors with core temperatures of $10^8\,\rm K$ similar to the \textsc{Arepo} donor model. While WD donors in realistic D$^6$ systems may in fact be much cooler than this, these models represent an optimistic assumption about the starting thermal state to which we can add more thermal energy due to supernova interaction. They are therefore a best case scenario for potentially achieving the thermally inflated state observed in D$^6$ runaway candidates.
Our white dwarf models are created using {\tt make\_co\_wd} in MESA. To create iso-thermal white dwarfs with temperature profiles close to the Arepo profile we turn off thermal neutrino cooling which would otherwise lead to white dwarfs with inverted interior temperature profiles due to thermal neutrino losses in the core.
We stop the cooling when the white dwarf reaches a central temperature of $10^8\mathrm{K}$. Our models consist of white dwarfs with final masses of 0.5 M$_{\odot}$, 0.64 M$_{\odot}$, 0.85 M$_{\odot}$, 0.95 M$_{\odot}$, and 1.1 M$_{\odot}$, with temperature profiles given in figure \ref{fig:wd_temp}. The main parameters of these models are given in \ref{table:params}.

\begin{table}[h!]
\caption{Relevant mass scales for the different models after the Nickel decay cut and for models without  helium tail}
\begin{center}
\begin{tabular}{ |c |c |c |c |c |}
\hline
Total Mass &$^{12}$C &$^{16}$O  & $^{4}$He & $^{56}$Ni \\
(M$_\odot$)& (M$_\odot$)& (M$_\odot$)&($10^{-2}$ M$_\odot$)&($10^{-2}$ M$_\odot$)\\
\hline
 0.50 & 0.247&0.245 & 0.21 &0.25 \\ 
 0.64 &0.316&0.314& 0.27& 0.32  \\  
 0.85 &0.420&0.416&0.35& 0.42 \\   
 0.95 &0.469&0.465&0.40& 0.47 \\    
 1.10 &0.543&0.539& 0.46& 0.55\\    
 \hline
\end{tabular}
\label{table:params}
\end{center}
\end{table}

All of these models are then relaxed to the composition profile taken directly from the \textsc{Arepo} remnant with a surface cut defined by the $0.64\,$M$_\odot$ before, scaled to be the same as a function of fractional mass coordinate $m/M$. Since the mass fractions are assumed to be the same, the total mass of individual elements is scaled by the mass of the model. Furthermore, to heat these models, we use the same $\Delta s_{\mathrm{Arepo}}$ and provide it as a heating term using Equation~\eqref{eq:heat3} over a time period of $0.01\,\mathrm{s}$. The maximum timestep is chosen to be $10^{-4}\,\mathrm{s}$, such that a minimum of 100 steps is needed to heat every model. All mixing and nuclear processes are turned off during this step. 
 
 This generalization of heating is approximate since we only have one simulation to rely on. For a less massive donor white dwarf, the binary separation before filling the roche-lobe will be greater, leading to a weaker shock. At the same time the less massive white dwarf is easier to perturb. The scaling is therefore non-trivial and not explored precisely here. For a more massive donor, the separation is smaller, but the internal pressure is higher. Analytic estimates using a white dwarf mass-radius relation indicate that the shock heating will be somewhat weaker for the most massive donors, though this relative shock strength scales relatively weakly with mass \citep{2019ApJ...887...68B}. So we believe that our method leads to a mild overestimate of the net shock heating for the most massive models, but as we will show, in the context of our work this only strengthens our result that shock heating alone is insufficient to explain the thermal state of the observed objects.

Since the mixing of the helium into the deeper layers is physically unrealistic, our main models only consist of white dwarfs relaxed to a composition profile where the helium in the inner 99\% of the white dwarf by mass coordinate is replaced with carbon. These models therefore only have surface helium, which is of the order $6\times10^{-5} M_\odot$. Discussion about models with the original \textsc{Arepo} composition is provided in the appendix \ref{B.Arepo}.

\section{Evolution for Observable Timescales}
\label{s.Evolution}

 We then carried out the subsequent evolution after relaxation/heating up to a maximum age of 100 Myr. We switched on nuclear burning, convection, and thermohaline mixing \citep{1980A&A....91..175K} (since the atmosphere has many heavy elements), as well as super-Eddington winds for mass loss of the energetically unbound material at the surface. The super-Eddington wind attempts to capture the mass loss that is driven by the 0.003 M$_\odot$ of $^{56}$Ni decay at earlier times when the luminosity of the star is greater than the Eddington luminosity. The decay energy of $^{56}$Ni deposits heat in excess of the binding energy of the outer layers, which tends to drive the envelope to a radiation-dominated super-Eddington state. The super-Eddington winds help provide a simple prescription for removing this mass from the model as a continuous wind until the heat deposited by Ni decay has either been lost or is no longer sufficient to unbind material. The super-Eddington wind in MESA has a mass loss rate that can be defined as a function of the luminosity $L$ and the local escape velocity $v_\mathrm{esc}$ as
 \begin{equation}
     \dot M=\frac{(L-L_{\mathrm{Edd}})}{\frac{1}{2} v^2_\mathrm{esc}}~.
 \end{equation}
 
 As the model reaches this luminosity, we apply extra pressure on the stellar atmosphere so that the model is able to evolve through the super-Eddington phase without struggling to resolve the atmospheric boundary condition in the radiation-dominated envelope. The pressure is removed after $10^3$ yrs and evolution continues normally. We therefore caution that the details of the observables for the early evolution (first $\sim$thousand years) should be interpreted as only very approximate, while the later details should generally be more reliable. Since we generally expect to compare to observed runaway remnants at least $10^4 - 10^6$ years after the supernova explosion, this should be sufficient for our purposes. The comparison of relaxation with heating during this phase of evolution is described in appendix \ref{A.Comparison}, while we focus on the main models here.

\subsection{Evolution of the 0.64 M$_\odot$ model}
\begin{figure}
\centering
\includegraphics[width=0.45\textwidth]
{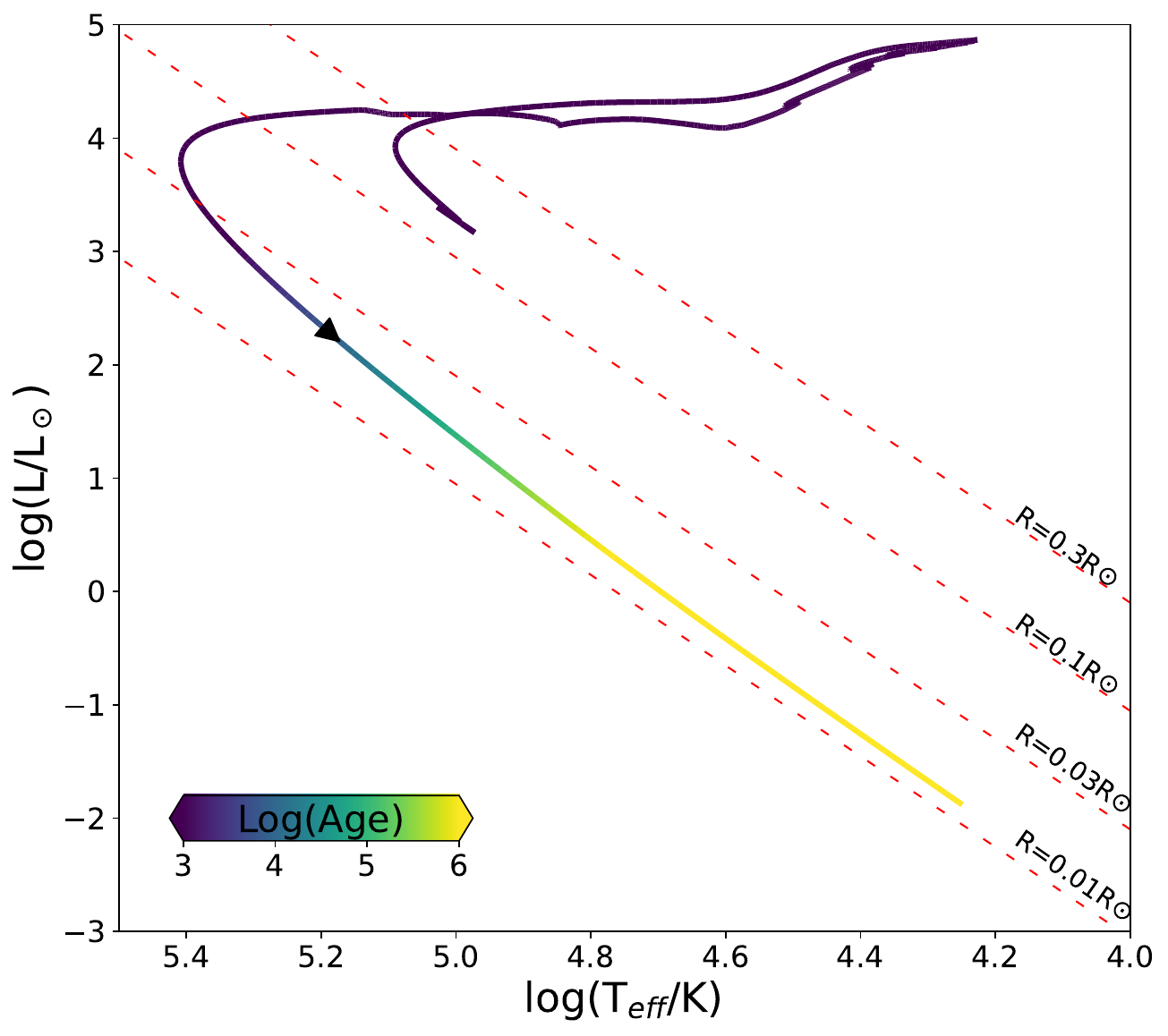}
\caption{Evolutionary track of the 0.64 M$_\odot$ model.}
\label{fig:def_evolve}
\end{figure}
 We first discuss the case of a $0.64$ M$_\odot$ model. This model has a composition similar to that of the \textsc{Arepo} remnant, but without any helium in the inner 99\% of the star by mass. The thermal profile of the model is based on the white dwarf evolution in MESA as shown in figure \ref{fig:wd_temp}.

\begin{figure}
\centering
\hspace*{-0.2cm}
\includegraphics[width=0.455\textwidth]{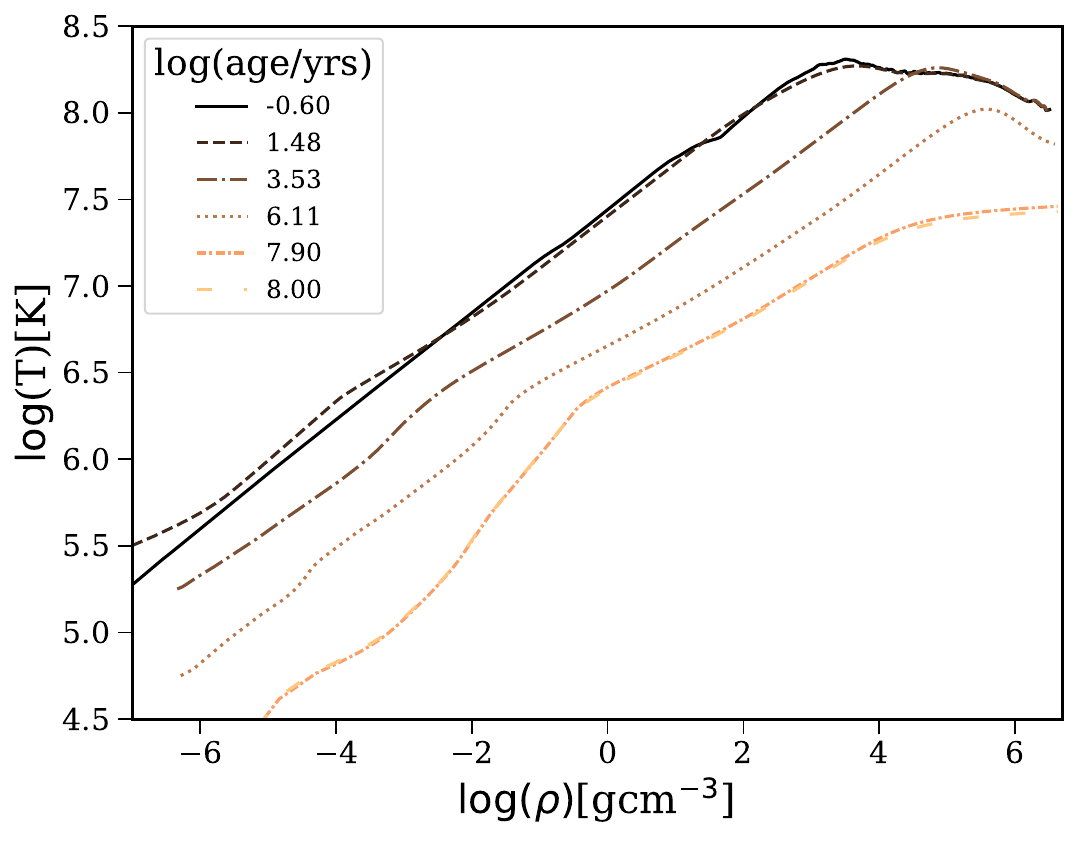} 
\caption{The structural profile of the $0.64M_\odot$ model as a function of its age. The puffing up and the later contraction of the star is clearly seen here. The core itself remains relatively unaffected and only the outer layers of the star are affected. }
\label{fig:rho_T_064_grid2}
\end{figure}

\begin{figure}
\centering
\hspace*{-0.2cm}
\includegraphics[width=0.45\textwidth]{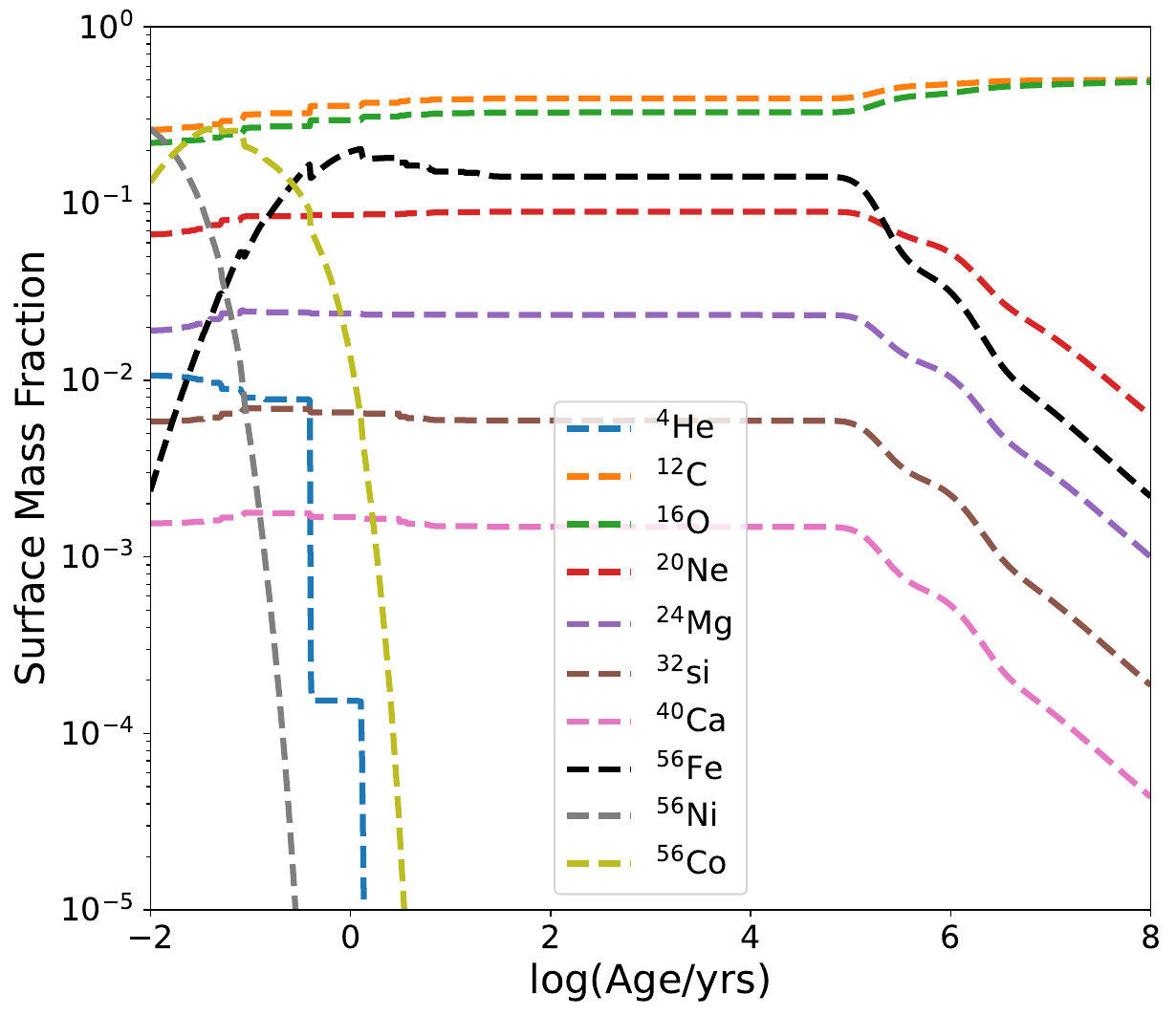} 
\caption{Mass fraction of elements which contribute significantly to the surface composition. The drop of these fractions in the later ages is due to thermohaline mixing.}
\label{fig:massfrac_064}
\end{figure}

The evolution can be broken down into a few major phases. The $\rho-T$ profile evolution is shown in figure \ref{fig:rho_T_064_grid2} for the $0.64\,$M$_\odot$ model. As the star is heated the temperature rises and it starts to become less dense. This heating is mostly restricted to the outer layers, and the core remains near the same structure as before. The peak temperature in the model lies near the surface where large amounts of heating occurred. Soon after the initial heating, the decay of $^{56}$Ni begins. Figure\, \ref{fig:massfrac_064} illustrates the evolution of the mass fractions of different elements at the surface (starting after 3 days). The $^{56}$Ni decay causes a super-Eddington wind which puffs up the star causing the density in the outer layers to drop. At the super-Eddington luminosity the star starts to lose mass. While the rest of $^{56}$Ni continues to decay, the daughter nuclei of $^{56}$Co starts decaying into $^{56}$Fe as well. As the half-life of $^{56}$Co is larger than $^{56}$Ni this process takes slightly longer. However, at the end of only a year of evolution, the surface nickel is gone and the iron content of the surface reaches a maximum. At this point, the star is at its maximum luminosity and radius, while the surface temperature is at a minimum. The net mass loss of this model, along with with the other models is shown in figure \ref{fig:massratio}.

Once all the $^{56}$Ni has decayed into $^{56}$Fe, the star readjusts thermally over a timescale of $100$ years. At this point, the mass loss rate starts to drop and the star begins to contract again. The surface elements, mainly those heavier than carbon and oxygen are mixed inside due to thermohaline mixing, and the surface eventually becomes C/O dominated.%
\footnote{Due to a limitation from the tabulated EOS used for the lowest density outer layers, the MESA thermohaline prescription does not recognise that thermohaline instability should occur for Fe lying on top of the C/O core (because the tabulation in terms of metallicity $Z$ does not recognise these as different). We therefore apply a minimal amount of mixing in the outer few percent of the mass of the star, which connects the Fe in these outer layers to the thermohaline mixing in the deeper core where the EOS resolves these differences, and therefore adequately captures the net effect of global interior thermohaline mixing pulling the heavy Fe away from the atmosphere.}
The absence of Helium is due to the fact that no Helium exists in our model for the inner 99\% by mass, and that the region where helium was retained is blown away by nickel-powered winds.

\subsection{Evolution of a range of masses}

\begin{figure}
\centering
\includegraphics[width=0.455\textwidth]{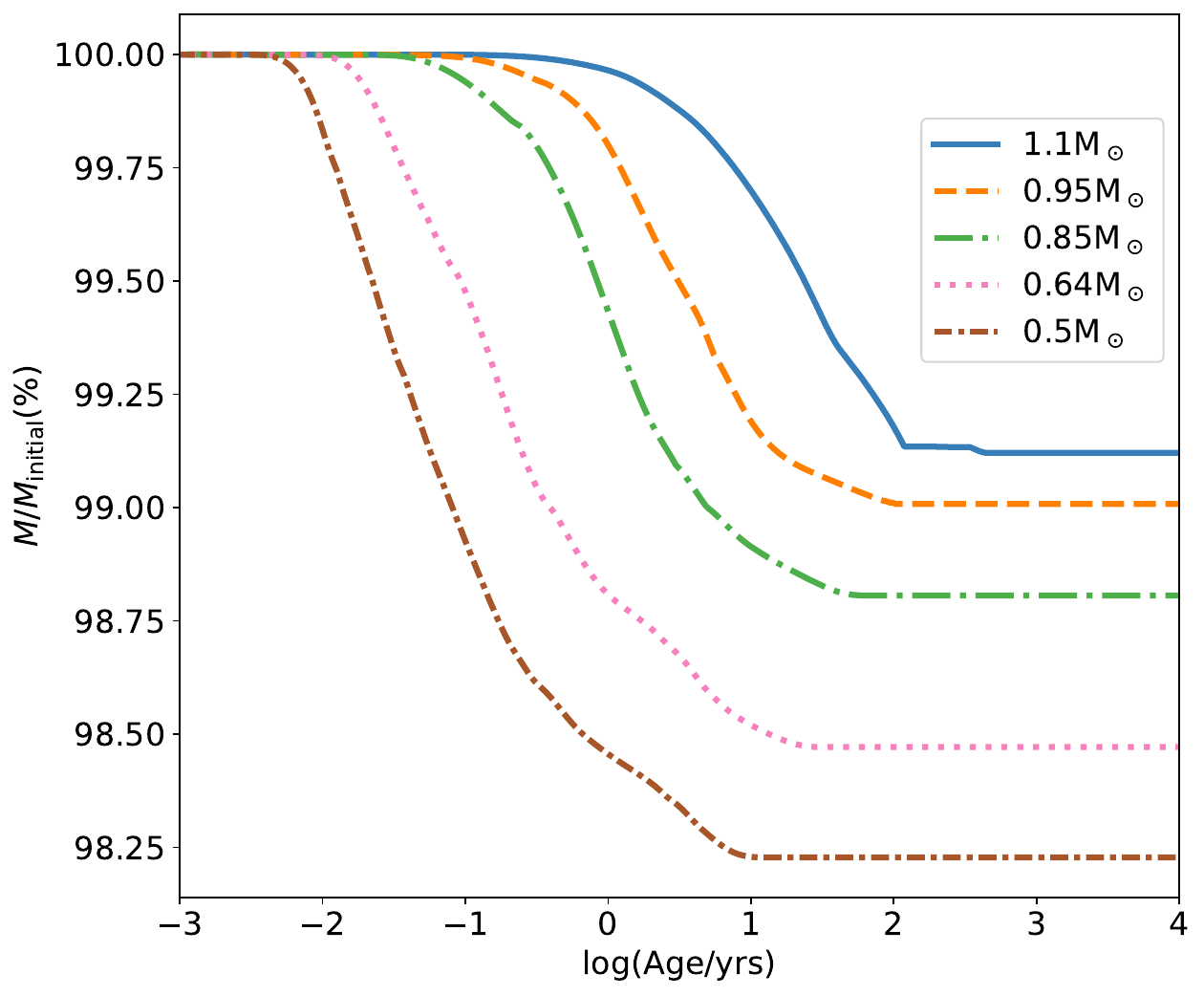} 
\caption{The ratio of mass of the star and its initial mass as a function of its age. There is no mass loss after $10^3\mathrm{yrs}$ for any model.}
\label{fig:massratio}
\end{figure}

The rest of the white dwarfs evolve in a similar way to the $0.64\,$M$_\odot$ model. The main difference is in the mass loss at the beginning of the evolution. Figure~\ref{fig:massratio} shows the amount of mass lost by these models over time due to super-Eddington winds. 
Overall, our models lose anywhere between $1-2\%$ of their initial mass. This mass loss happens in the layers which were heavily contaminated by $^{56}$Ni. Due to this mass loss the total mass fraction of $^{56}$Fe at the end of the evolution is lower than would be expected if all the $^{56}$Ni was allowed to decay without any mass loss. 

\begin{figure}
\centering
\hspace*{-0.2cm}
\includegraphics[width=0.45\textwidth]{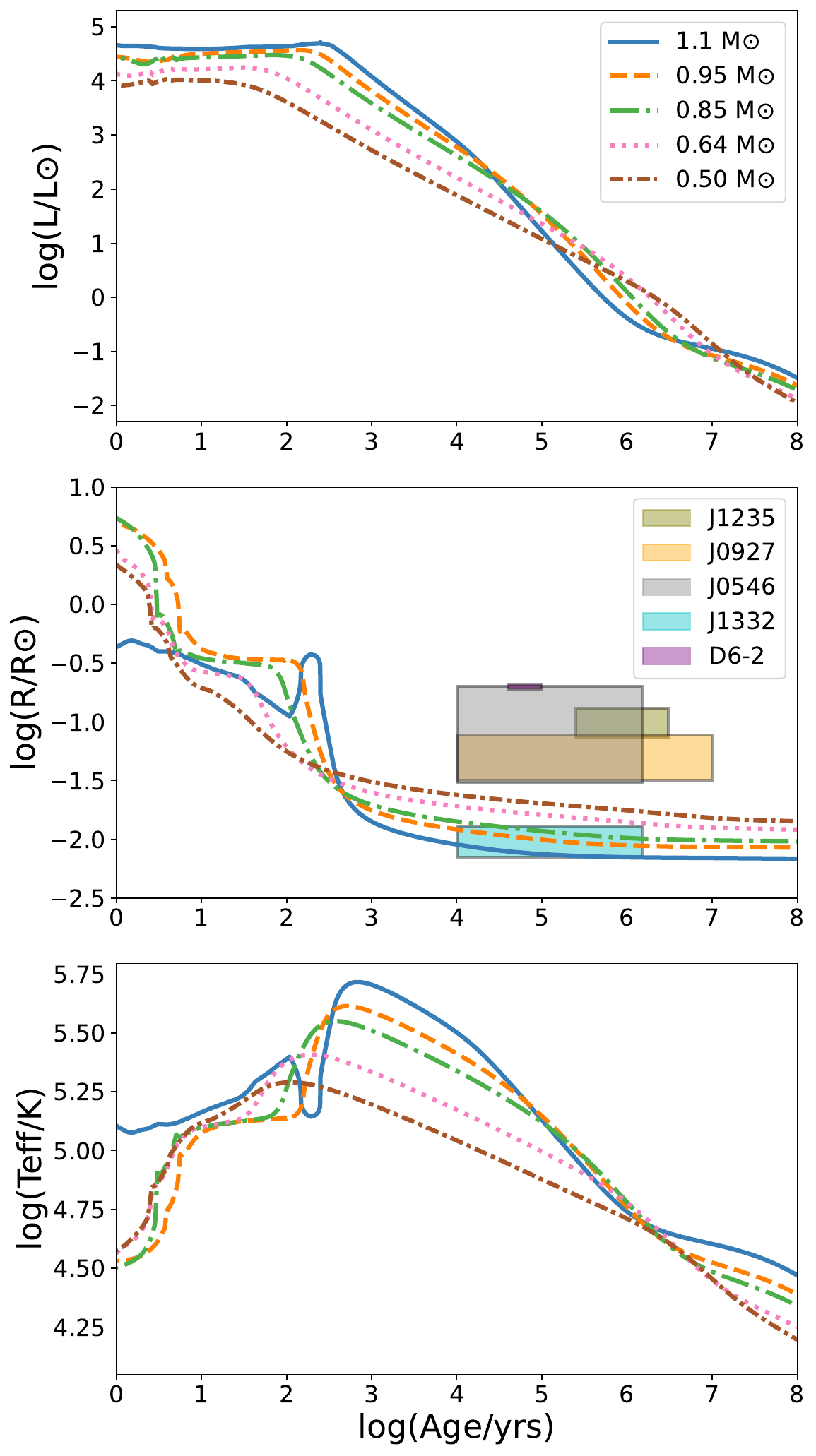} 
\caption{Evolution of the surface luminosity, radius, and temperature of our models. D6 observations are plotted here for the radius. While the radii of D6-1 and D6-3 are similar to those of D6-2, their age estimate is less certain and spans a bigger region. They are not plotted here so as to not overcrowd the diagram. The radius bump at $10^2$ yr for the $1.1\, M_\odot$ is due to a lowering of the extra surface pressure (applied to the early evolution for numerical convergence), and is non-physical. The $T_{\rm eff}$ drop corresponds to this.}
\label{fig:lrt_2}
\end{figure}

The evolution of surface luminosity, radius, and effective temperature are shown in Figure \ref{fig:lrt_2}. We also overplot the recent D$^6$ observations. We estimate ages of the observations by assuming that the SN ejection should most likely come from somewhere in the Galactic disk. The maximum ages of J1332, J0546, and J1235 are estimated using $t=(z+1)/v_z$, where $z$ is there present height above the galactic plane, $v_z$ is the velocity in z-direction, and 1 kpc is added as an estimate of the scale height of the thick disk (\citealt{2017ApJ...850...25L}). For D6-2 we use the minimum and maximum ages of the supernova remnant associated with it (\citealt{Shen2018a}). As J0927 is coming towards us, we use minimum and maximum ages of $10^4$ and $10^7$ years respectively. 

\citet{2024A&A...682A..42W} studied and used NLTE models for spectral fitting of the hottest stars from \citet{2023arXiv230603914E}. However, they used cooling tracks of post-AGB stars to get other parameters like the mass, radius, and luminosity. Since, the D$^6$ stars do not fit into this evolutionary channel, we used the radii from \citet{2023arXiv230603914E} instead, but base our uncertainties taking the previous work of \citet{2024A&A...682A..42W} into account. For J332, which is the star closest to the white dwarf cooling track, we used the gravity from \citet{2024A&A...682A..42W}, and calculate the uncertainties in the radius assuming a mass distribution from $0.5-1.1$ M$_\odot$. We adapted the parameters for the original 3 cool D$^6$ candidate stars  from the work of \citet{2022MNRAS.512.6122C} and \citet{2018ApJ...865...15S}.

This figure again shows the initial puffing up due to the heating and the subsequent nickel decay. However, like before, this mass loss only lasts for a few hundreds of years, after which only iron is left behind. The models reach the cooling track quickly. As their luminosities are relatively high at a radii of $0.01$ R$_\odot$, they are also quite hot ($>10^5 \mathrm{K}$). Only J1332 is consistent with our models. Every other observation lies outside of our evolutionary tracks due to their inflated nature.

\begin{figure*}
\centering
\includegraphics[width=0.455\textwidth]{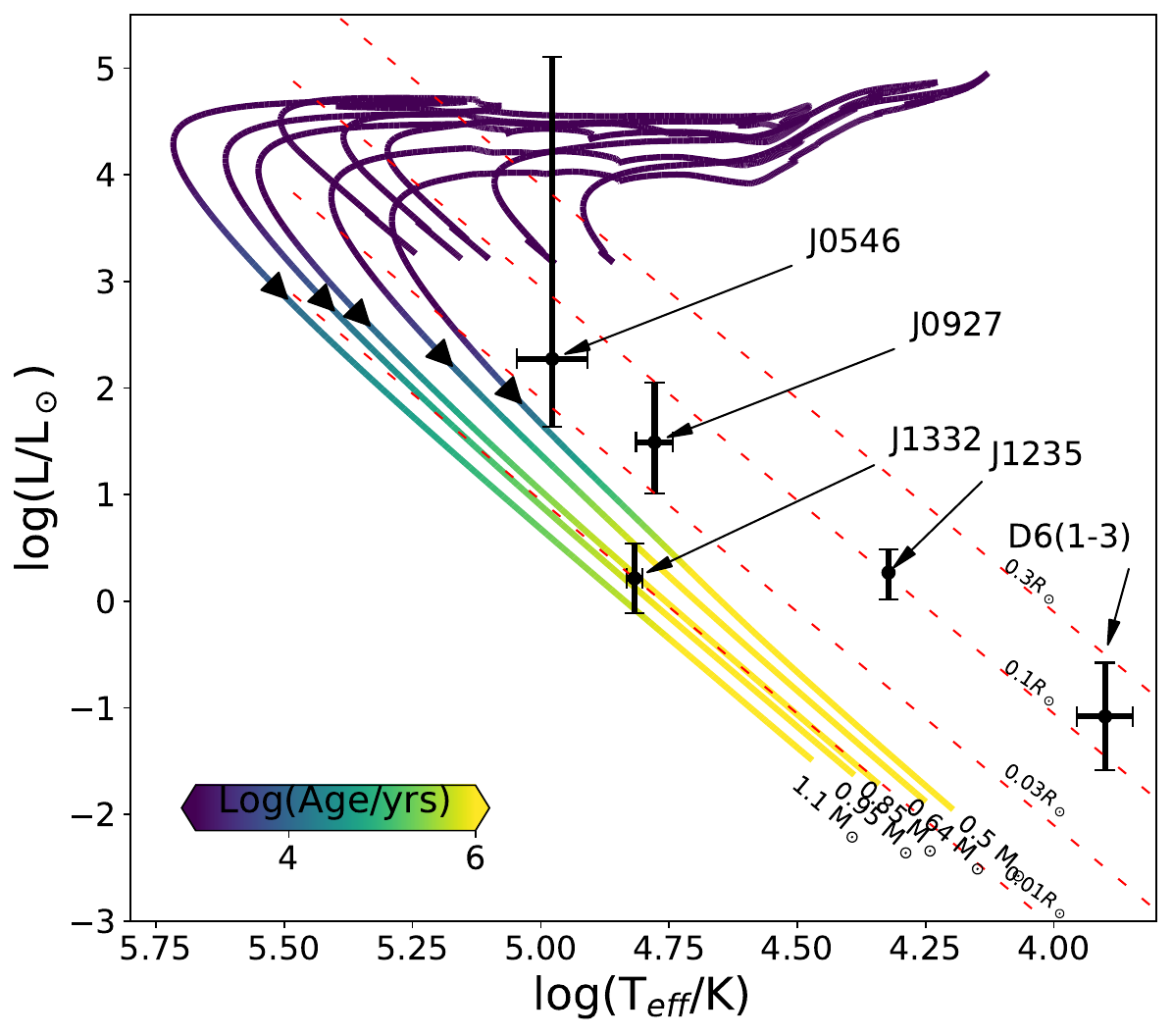} 
\includegraphics[width=0.45\textwidth]{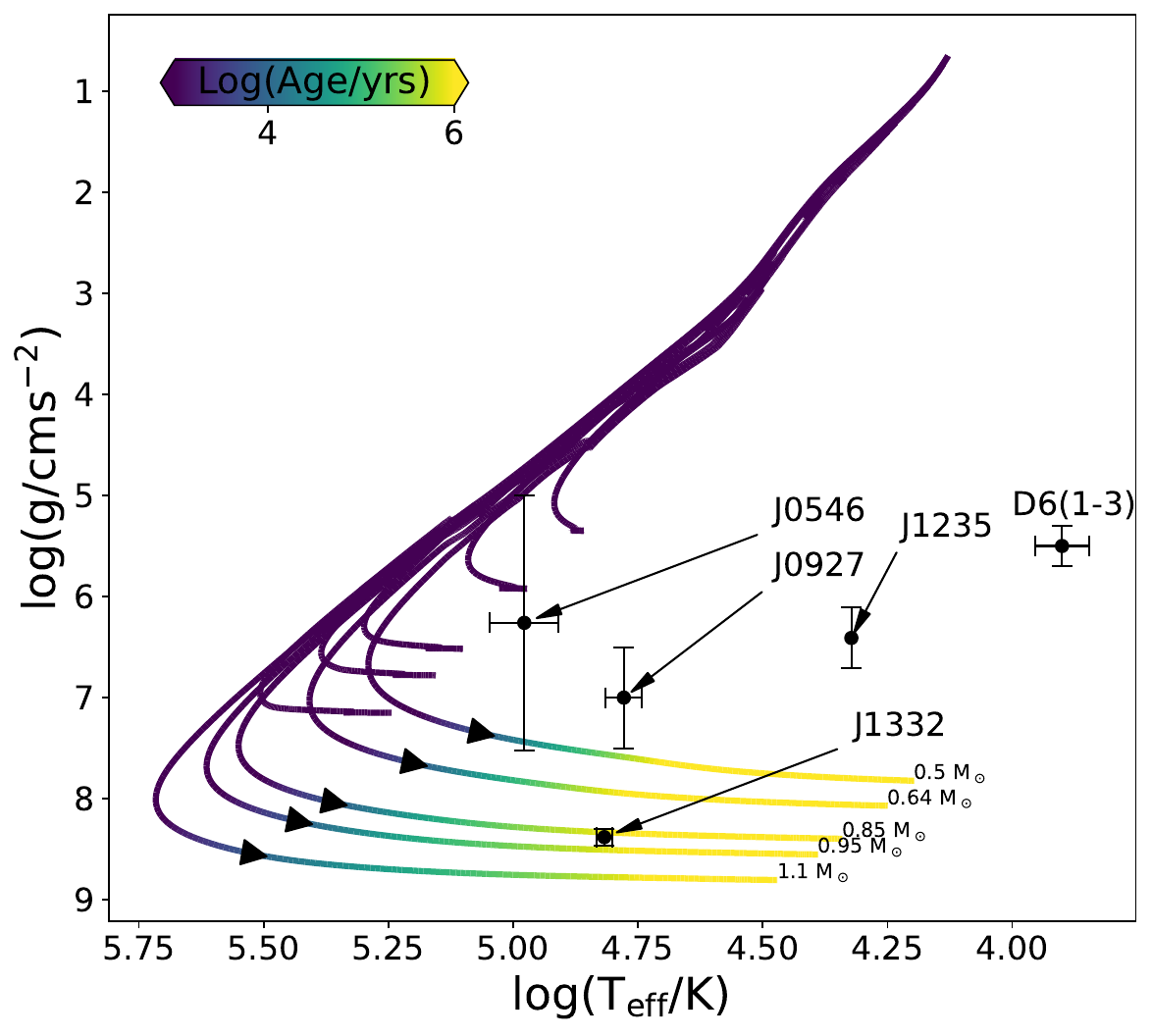}
\caption{\textbf{[left panel]} HR diagram for all WD models. All observed D$^6$ stars are plotted as black points. Dashed red lines represent lines of constant radii. Colorbar shows the log(age) of the star. The  arrows mark the points when the model is $10^4$ yr old. \textbf{[right panel]} The Kiel diagram for the same model. The high uncertainty in J0546 comes from combining the minimum and maximum values from \citet{2024A&A...682A..42W} and \citet{2023arXiv230603914E}, which have different ways of measuring the radius that lead to inconsistent results.}
\label{fig:evolve_tailless}
\end{figure*}

\section{Discussion of results and comparison with observations}
 We compare our models to observed D$^6$ runaway candidates in this section.
\subsection{Evolutionary diagrams}
The HR and Kiel diagrams of all the models are shown in Figure \ref{fig:evolve_tailless}, with the D$^6$ observations over-plotted. We took the temperatures  and gravities of the three hottest stars from \citet{2024A&A...682A..42W}, from \citet{2023arXiv230603914E} for J1235, and from \citet{2022MNRAS.512.6122C} for the cooler D6-1,2,3. We calculated the luminosity using:

\begin{equation}
(L/L_\odot) = (R/R_\odot)^2\left(\frac{T_{\text{eff}}}{5800\,\rm K}\right)^4~.
\label{eq:bb}
\end{equation}

The Kiel diagram shows clearly that none of the stars, except  J1332, can be explained by our models. For the case of J1332, our models suggest that the star is consistent with being on the cooling track. This was also reported by \citet{2024A&A...682A..42W}, who classified this star as a DA-white dwarf. Our result is also in agreement with the mid-plane age (time taken by the star to reach its present height above plane assuming ejection in the disk at $z = 0$ kpc) of this star calculated by \citet{2023arXiv230603914E} of around 0.6 Myr. For J0546, whose mid-plane age is also similar, the uncertainties are big enough that no conclusion can be reached. The slight overlap of this star with our tracks is possibly due to an overestimation of the uncertainties. That the lower limit of the gravity of the star is overestimated can be seen by the fact that if the star were younger than $10^3\, \mathrm{yrs}$, a supernova remnant would be observable. Furthermore, such a supernova event would have been seen within the recorded history of human civilisation. As no such record exists, it is highly unlikely that the star is so young as a runaway.
\subsection{Masses and radii}

One of the main aims of this study was to map the \textsc{Arepo} profile of the surviving white dwarf from our single explosion simulation to multiple donor masses. The subsequent evolution for all models showed similar evolutionary tracks, with only the time spent in the super-Eddington luminosity regime being different. 

 Despite starting with relatively hot and luminous white dwarfs in our models, which is inconsistent with searches for runaway stars in supernova remnants \citep{2013ApJ...774...99K,2014ApJ...782...27K, 2018MNRAS.479..192K, 2022ApJ...933L..31S, 2023ApJ...950L..10S}, our models do not puff up to match recent observations. Our models all tend toward radii on the order of $0.01\, $R$_\odot$ within a few thousand years, where their high luminosities  lead to the high surface temperatures of our stars ($>10^5$ K). While temperatures close to this have been observed for the hottest D$^6$ stars, many of the observed D$^6$ stars instead have much larger radii and cooler surface temperatures ($6000-7000$ K), which our models are currently unable to explain. As shown by the radial evolution in the middle panel of Figure~\ref{fig:lrt_2}, only 1 star, J1332 is close to the predicted stellar radius.

Table~\ref{table:1} shows the estimated orbital velocities that a donor would have for our models and Table~\ref{table:2} shows the estimated masses for the observed stars based on observed velocities. For D6-1, D6-3, J1235, J0927 the estimated masses are near or above $1\,$M$_\odot$. For  all of the stars except D6-2, the maximum mass within $1\sigma$ uncertainties is more than 1.1 M$_\odot$, and especially for the fastest ones is even greater than 1.2 M$_\odot$. \cite{Shen2024} recently pointed out that donor stars less massive than about 1.0\,$M_\odot$ may also support double detonations that could destroy the donor in addition to the accretor and produce no runaway, while more massive donors may be more likely to survive and produce a runaway. This may offer an explanation for why the observed candidates tend to have runaway velocities well above 2000\,km\,s$^{-1}$. At the same time, our 1.1 M$_\odot$ model has the most compact radius and is the farthest away from the observations. J1332 might be explained by our models greater than or equal to 0.85 M$_\odot$, while J0546 cannot be completely ruled out.

\begin{table}[h!]
\caption{Approximate orbital velocities of donors before the supernova.}
\begin{center}
\begin{tabular}{ |c |c |}
\hline
Mass (M$_\odot$) & Orbital velocity (km\,s$^{-1}$)\\
\hline
 0.50 & 1700  \\ 
 0.64 & 1800  \\  
 0.85 & 2000 \\   
 0.95 & 2300 \\    
 1.10 & 2500 \\    
 \hline
\end{tabular}
\tablefoot{ Values taken from Figure~4 of \citet{2021ApJ...923L..34B} }
\label{table:1}
\end{center}
\end{table}

\begin{table}[h!]
\caption{Ejection velocities with $1\sigma$ uncertainties of D$^6$ candidate runaways}
\begin{center}
\begin{tabular}{ |c |c | c |c |}
\hline
Star & V$_\textrm{ejection}$ (km\,s$^{-1}$)&  M$^{\mathrm{min}}_{\mathrm{donor}}$ (M$_\odot$)& M$^{\mathrm{est}}_{\mathrm{donor}}$ (M$_\odot$)\\
\hline
 D6-1 & $2254^{+248}_{-185}$ &0.8& 1.0 \\ 
 D6-2 & $1051^{+62}_{-54}$&0.1&0.1\\
 D6-3 & $2393^{+377}_{-391}$& 0.7&1.1\\  
 J1235 & $2471^{+351}_{-345}$ & 0.8&1.1 \\    
 J0927 & $2519^{+271}_{-147}$ &1.0&1.1\\    
 J0546 & $1864^{+682}_{-416}$ & 0.2&0.7\\
 J1332 & $1619^{+707}_{-320}$ & 0.2&0.4\\
 \hline
\end{tabular}

\tablefoot{Values taken from \citet{2023arXiv230603914E}. The minimum donor mass and estimated donor masses are shown based on the minimum ejection velocities and median ejection velocity using \citet{2021ApJ...923L..34B} to calculate the masses. }
\label{table:2}
\end{center}
\end{table}
\subsection{Thermal Time and Shock}
\begin{figure}
\centering
\includegraphics[width=0.45\textwidth]{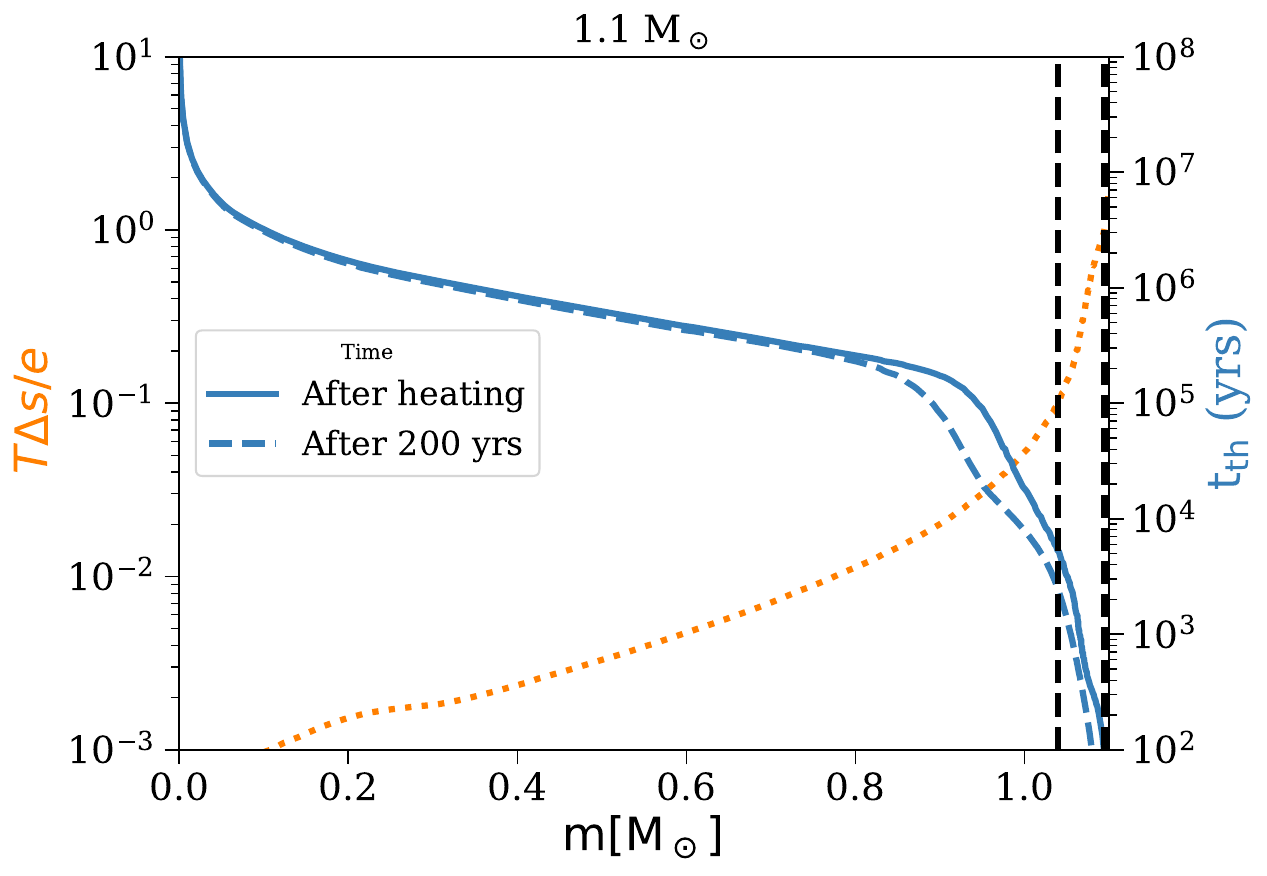} 
\includegraphics[width=0.45\textwidth]{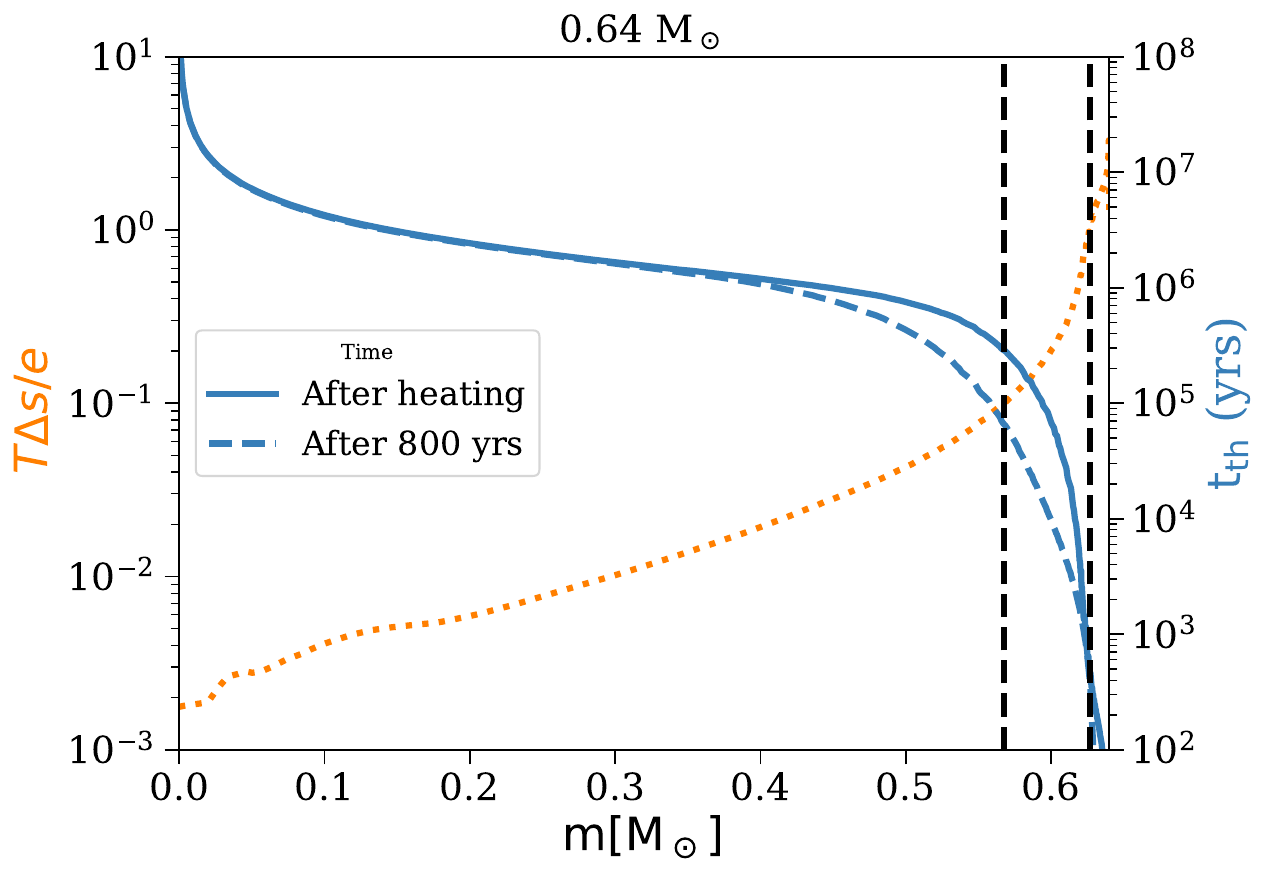} 
\caption{Fraction of input heat and internal energy (left axis, orange), and thermal time (right axis, blue) of $1.1\,$M$_\odot$ \textbf{[top]} and $0.64\,$M$_\odot$ \textbf{[bottom]} models. Dashed and solid black lines represent 10\% and 100\% of internal energy input as shock heating respectively. The thermal time is plotted for ages of the order $0$ and $>10^2$ years. }
\label{fig:tds}
\end{figure}

One of the main results of our work is to show that a supernova shock combined with a deposition of heavy elements like nickel is not enough to explain the parameters of the observed D$^6$ white dwarfs with our current understanding of stellar physics. We show this directly in Figure~\ref{fig:tds}. We plot the ratio of input heat and the internal energy of the white dwarf as a function of the mass coordinate for 0.64 M$_\odot$ and 1.1 M$_\odot$ models. The input heat or $T\Delta s$ is calculated using the difference of entropy of the initial white dwarf and the post-heating white dwarf from \textsc{Arepo}, multiplied by the initial temperature profile of the white dwarf. The internal energy $e$ is the internal energy of the unperturbed white dwarf. The ratio is therefore a metric to highlight the fractional heating that has occurred in the white dwarf model relative to its original structure. The second axis shows the thermal time of these layers, which is the local thermal diffusion timescale, given in Equation~\eqref{eq:td}. We plot the thermal time at different periods of evolution (after heating and $\sim10^2-10^3$ years later after mass loss occurred).

The plot shows clearly that the input heat due to the supernova shock is too low to significantly perturb the structure and inflate the star in layers of the white dwarf where the thermal time is of the order of lifetime of the observed D6 white dwarfs. Furthermore, some of the outer layers are eventually stripped away as a result of nickel-powered mass loss. As the white dwarf loses these layers, it decompresses and the thermal time of underlying layers decreases, especially near the surface.

\subsection{Opacities, structure, and surface abundances}

We have shown above that the shock strength is not enough to perturb the inner regions of a white dwarf and that the thermal time is too short in regions where this shock is high enough. One key limitation of this modelling is that we do not take into account some physics that could influence the precise surface compositions of these runaways at evolutionary times greater than a few thousand years. Thermohaline mixing causes heavy elements to mix deeper into the model and away from the photosphere (see Figure~\ref{fig:massfrac_064}), and our modelling does not account for other physics such as element diffusion and radiative levitation that could selectively influence the relative compositions at the photosphere. The resulting surface compositions may impact opacities by orders of magnitude when heavier elements are involved. A significant proportion of iron on the surface, for instance, could lead to unusual opacities. As the radius of the star is dependent on the opacity of different layers, a more precise abundance and opacity calculation is needed. Most candidate spectra show an abundance of lines from heavy elements including C, O, Ne, and Fe, though most have not yet been analysed for detailed surface abundances \citep{2018ApJ...865...15S,2022MNRAS.512.6122C,2023arXiv230603914E}. On the other hand, \citet{2024A&A...682A..42W} found that J1332 has a high surface hydrogen and helium abundance. They speculate that this might be the result of J1332 passing through a molecular cloud. Our models show no surface hydrogen or helium on observable timescales, but a conclusive statement can only be made after the exact surface mixing processes are taken into account.

\begin{figure}
\centering
\includegraphics[width=0.45\textwidth]{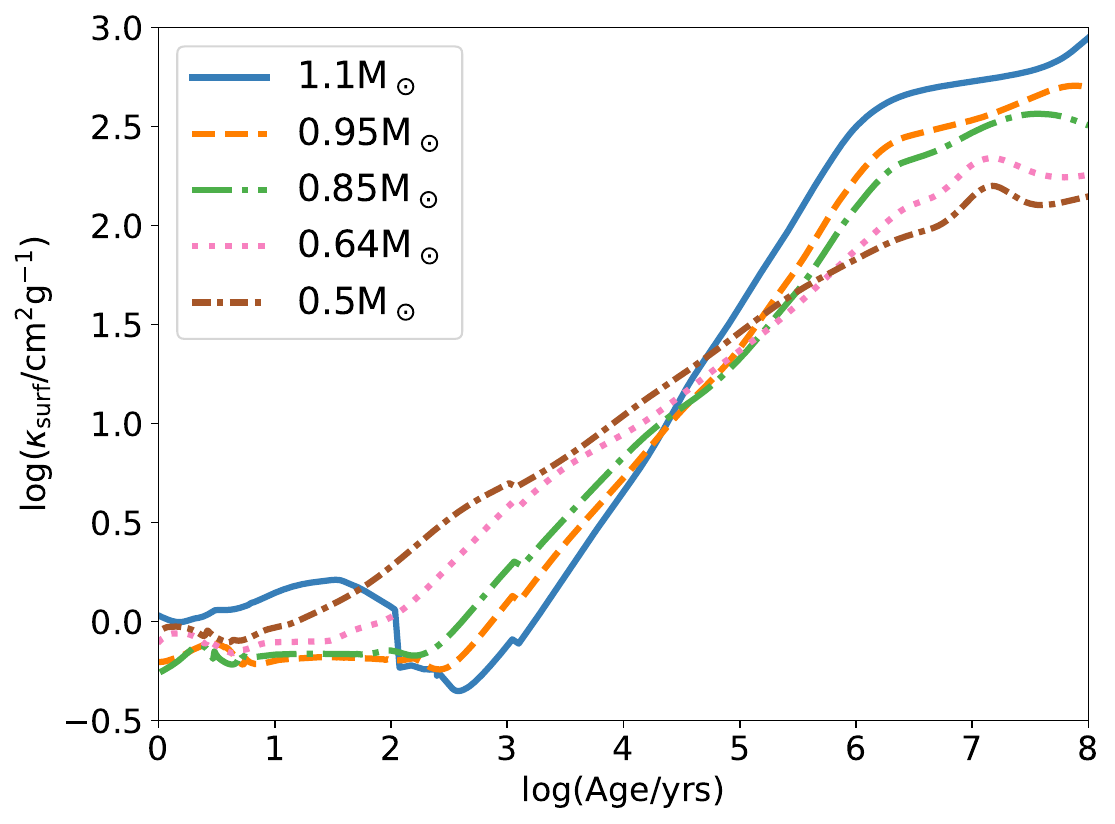} 
\caption{The surface opacity of the star as a function of its age. Surface opacities increase by orders of magnitude as the star cools down. the slightly different behaviour of the 1.1 M$_\odot$ in the beginning is due to the extra pressure applied to evolve during the super-Eddington phase.}
\label{fig:surfop}
\end{figure}

The default MESA opacities are in the form of tables for a given set of composition, temperature, and density from OPAL tables \citep{1993ApJ...412..752I}, which account for C/O enrichment for high $Z$ compositions assumed to be ashes of He burning. For our WD models, the surface is polluted with heavier elements due to the supernova ejecta, and no pre-computed opacity tables are available that accurately reflect the unusual surface composition of our initial models. However, there are no noticeable differences between the tabulated opacities and our model opacities at times more than a few 1000 yrs because the surface has again become dominated by C/O due to mixing.
The surface opacity rises as the effective temperature drops, as seen in Figure~\ref{fig:surfop}. As the white dwarf gets older and cools back down the carbon, oxygen, and iron opacities increase. 
A model which better captures surface mixing, and includes a better model of diffusion and radiative levitation might have a different composition at the surface at later times than what we find. A proper study of these mixing processes was outside of the scope of this work, but is important for any future studies on this topic.

\section{Conclusions}

In this paper we have presented results for the first time of 1D stellar evolution of a D$^6$ runaway star produced from a simulation of the D$^6$ supernova detonation scenario. We used composition and temperature-density profiles of the white dwarf from previously published hydrodynamical simulations of WD binary supernovae made in \textsc{Arepo}. We were able to map these into MESA and heat white dwarfs representing donors in the mass range $0.5\,$M$_\odot - 1.1\,$M$_\odot$. We then evolved these models to 100 Myr after the supernova explosion.

Our work clearly shows that with our current understanding of the physics of stellar structure for these models, the post-supernova explosion shock is not enough to explain runaway observations. Our WD donor models have very high interior temperatures of $10^8$\,K, which would correspond to a cooling age of no more than a few Myr for normal WDs, representing a maximally optimistic choice for their starting thermal state. Despite this the white dwarfs do not heat and expand for a timescale which matches the observed inflated D$^6$ stars. The only star which lies on our evolutionary tracks is J1332, which has possibly been a runaway long enough to return back to the white dwarf cooling track. Our results show that the stars spend only few hundred years puffed up to around 0.1 R$_\odot$, powered by the decay of $^{56}$Ni and shock heating of their outer layers. The significant radioactive heating at the surface from nickel decay leads to most of the accreted supernova ejecta being quickly lost again as a wind, and does not appear to produce heating that can penetrate deeper into the core. After all the nickel has been converted to iron and after the luminosity returns back to sub-Eddington, the white dwarf starts to cool down, but always having reached a compact structure on the WD cooling track for our models.

This modelling contrasts with previous more successful attempts to explain inflated states of runaways from hot subdwarf donors \citep{2019ApJ...887...68B} or bound Iax remnants \citep{2019ApJ...872...29Z}, for which some models are able to reach unusual inflated states for Myr timescales. However, only a WD donor can explain the observed velocities for many of the hypervelocity runaways now being discovered. Unlike the hot subdwarf or Iax scenarios, in the case of WD donors in D$^6$ binaries, the central pressure of the donor WD is much greater compared to the shock driven by SN interaction. Therefore the shock that traverses the core of the donor is relatively weak and unable to significantly perturb its thermal structure. Significant heating is largely confined to superficial layers with short thermal timescales near the surface, and these layers are either quickly lost through winds or able to thermally readjust and settle back into a compact configuration over timescales much shorter than Myr in our models. Explaining the unusually inflated states of many of the observed high velocity runaways requires either (1) an additional source of significant heating or (2) additional physics influencing the structure and evolution that is not currently accounted for in our modelling.

While our work has focused on the long-term evolution of the structure of the white dwarf, the next step in refining this modelling will be precise modelling of the surface abundances. Our thermohaline mixing prescription shows that the heavier elements sink down into the white dwarf such that surface abundance of these stars eventually becomes dominated by carbon and oxygen dredged from the core. A better prescription of diffusion and radiative levitation which acts on the heavier elements like iron and changes surface composition should be investigated in the future. Furthermore, 3 dimensional hydrodynamical simulations dealing with the most massive donors should also be investigated for a better understanding of the post-supernova shock heating to improve upon our current approximate scaling based on just one detailed simulation of a $0.7\,M_\odot$ WD donor.
\section{Data Availability}
The MESA input files and inlists for relaxation, heating, and evolution can be accessed at \url{https://doi.org/10.5281/zenodo.13477859}

\begin{acknowledgements}
This project was originally started as part of the Kavli Summer Program which took place in the Max Planck Institute for Astrophysics in Garching in July 2023, supported by the Kavli Foundation. We are grateful to Stephen Justham, Selma de Mink, 
 and Jim Fuller for enriching discussions. We would like to thank the anonymous referee for their helpful report. A.B. was supported by the Deutsche Forschungsgemeinschaft (DFG) through grant GE2506/18-1.
  K.J.S.\ was supported by NASA through the Astrophysics Theory Program (80NSSC20K0544) and by NASA/ESA Hubble Space Telescope programs \#15871 and \#15918.

  W.E.K. was supported by NSF Grants OAC-2311323, AST-2206523, and NASA/ESA HST-AR-Theory HST-AR-16613.002-A. K.E. was supported in part by HST-GO-17441.001-A. AB and ASR would like to thank Rob Farmer for his support with PyMESA.
\end{acknowledgements}
\section*{Contributor Roles}
\begin{itemize}
    \item Conceptualisation: Evan Bauer, Ruediger Pakmor, Ken Shen
    \item Data curation: Aakash Bhat, Ruediger Pakmor
    \item Formal analysis: Aakash Bhat, Evan Bauer
    \item Funding acquisition: Stephen Justham, Selma de Mink, Stephan Geier, Evan Bauer
    \item Investigation: Aakash Bhat
    \item Methodology: Aakash Bhat, Evan Bauer, Abinaya Swaruba Rajamuthukumar
    \item Supervision: Evan Bauer, Ruediger Pakmor, Ilaria Caiazzo
    \item Visualisation: Aakash Bhat, Ruediger Pakmor
    \item Writing -- original draft: Aakash Bhat, Evan Bauer, Ruediger Pakmor, Ken Shen
    \item Writing -- review \& editing: Evan Bauer, Ruediger Pakmor, Ken Shen, Ilaria Caiazzo, Abinaya Swaruba Rajamuthukumar, Kareem El-Badry, Wolfgang Kerzendorf 
\end{itemize}
\bibliographystyle{aa}
\bibliography{example}

\begin{thebibliography}{65}
\expandafter\ifx\csname natexlab\endcsname\relax\def\natexlab#1{#1}\fi

\bibitem[{{Angulo} {et~al.}(1999){Angulo}, {Arnould}, {Rayet}, {Descouvemont},
  {Baye}, {Leclercq-Willain}, {Coc}, {Barhoumi}, {Aguer}, {Rolfs}, {Kunz},
  {Hammer}, {Mayer}, {Paradellis}, {Kossionides}, {Chronidou}, {Spyrou},
  {degl'Innocenti}, {Fiorentini}, {Ricci}, {Zavatarelli}, {Providencia},
  {Wolters}, {Soares}, {Grama}, {Rahighi}, {Shotter}, \& {Lamehi
  Rachti}}]{Angulo1999}
{Angulo}, C., {Arnould}, M., {Rayet}, M., {et~al.} 1999, \nphysa, 656, 3

\bibitem[{{Bauer} {et~al.}(2021){Bauer}, {Chandra}, {Shen}, \&
  {Hermes}}]{2021ApJ...923L..34B}
{Bauer}, E.~B., {Chandra}, V., {Shen}, K.~J., \& {Hermes}, J.~J. 2021, \apjl,
  923, L34

\bibitem[{{Bauer} {et~al.}(2019){Bauer}, {White}, \&
  {Bildsten}}]{2019ApJ...887...68B}
{Bauer}, E.~B., {White}, C.~J., \& {Bildsten}, L. 2019, \apj, 887, 68

\bibitem[{{Boos} {et~al.}(2024){Boos}, {Townsley}, \& {Shen}}]{boos2024}
{Boos}, S.~J., {Townsley}, D.~M., \& {Shen}, K.~J. 2024, \apj, 972, 200

\bibitem[{{Boos} {et~al.}(2021){Boos}, {Townsley}, {Shen}, {Caldwell}, \&
  {Miles}}]{2021ApJ...919..126B}
{Boos}, S.~J., {Townsley}, D.~M., {Shen}, K.~J., {Caldwell}, S., \& {Miles},
  B.~J. 2021, \apj, 919, 126

\bibitem[{{Cassisi} {et~al.}(2007){Cassisi}, {Potekhin}, {Pietrinferni},
  {Catelan}, \& {Salaris}}]{Cassisi2007}
{Cassisi}, S., {Potekhin}, A.~Y., {Pietrinferni}, A., {Catelan}, M., \&
  {Salaris}, M. 2007, \apj, 661, 1094

\bibitem[{{Chandra} {et~al.}(2022){Chandra}, {Hwang}, {Zakamska}, {Blouin},
  {Swan}, {Marsh}, {Shen}, {G{\"a}nsicke}, {Hermes}, {Putterman}, {Bauer},
  {Petrosky}, {Dhillon}, {Littlefair}, \& {Ashley}}]{2022MNRAS.512.6122C}
{Chandra}, V., {Hwang}, H.-C., {Zakamska}, N.~L., {et~al.} 2022, \mnras, 512,
  6122

\bibitem[{{Chugunov} {et~al.}(2007){Chugunov}, {Dewitt}, \&
  {Yakovlev}}]{Chugunov2007}
{Chugunov}, A.~I., {Dewitt}, H.~E., \& {Yakovlev}, D.~G. 2007, \prd, 76, 025028

\bibitem[{{Cyburt} {et~al.}(2010){Cyburt}, {Amthor}, {Ferguson}, {Meisel},
  {Smith}, {Warren}, {Heger}, {Hoffman}, {Rauscher}, {Sakharuk}, {Schatz},
  {Thielemann}, \& {Wiescher}}]{Cyburt2010}
{Cyburt}, R.~H., {Amthor}, A.~M., {Ferguson}, R., {et~al.} 2010, \apjs, 189,
  240

\bibitem[{{Dan} {et~al.}(2011){Dan}, {Rosswog}, {Guillochon}, \&
  {Ramirez-Ruiz}}]{Dan2011}
{Dan}, M., {Rosswog}, S., {Guillochon}, J., \& {Ramirez-Ruiz}, E. 2011, \apj,
  737, 89

\bibitem[{{Dan} {et~al.}(2012){Dan}, {Rosswog}, {Guillochon}, \&
  {Ramirez-Ruiz}}]{Dan2012}
{Dan}, M., {Rosswog}, S., {Guillochon}, J., \& {Ramirez-Ruiz}, E. 2012, \mnras,
  422, 2417

\bibitem[{{El-Badry} {et~al.}(2023){El-Badry}, {Shen}, {Chandra}, {Bauer},
  {Fuller}, {Strader}, {Chomiuk}, {Naidu}, {Caiazzo}, {Rodriguez}, {Nagarajan},
  {Yamaguchi}, {Vanderbosch}, {Roulston}, {G{\"a}nsicke}, {Han}, {Burdge},
  {Filippenko}, {Brink}, \& {Zheng}}]{2023arXiv230603914E}
{El-Badry}, K., {Shen}, K.~J., {Chandra}, V., {et~al.} 2023, The Open Journal
  of Astrophysics, 6, 28

\bibitem[{Farmer \& Bauer(2018)}]{pymesa}
Farmer, R. \& Bauer, E.~B. 2018, pyMesa

\bibitem[{{Ferguson} {et~al.}(2005){Ferguson}, {Alexander}, {Allard}, {Barman},
  {Bodnarik}, {Hauschildt}, {Heffner-Wong}, \& {Tamanai}}]{Ferguson2005}
{Ferguson}, J.~W., {Alexander}, D.~R., {Allard}, F., {et~al.} 2005, \apj, 623,
  585

\bibitem[{{Fink} {et~al.}(2007){Fink}, {Hillebrandt}, \&
  {R{\"o}pke}}]{2007A&A...476.1133F}
{Fink}, M., {Hillebrandt}, W., \& {R{\"o}pke}, F.~K. 2007, \aap, 476, 1133

\bibitem[{{Fuller} {et~al.}(1985){Fuller}, {Fowler}, \& {Newman}}]{Fuller1985}
{Fuller}, G.~M., {Fowler}, W.~A., \& {Newman}, M.~J. 1985, \apj, 293, 1

\bibitem[{{Geier} {et~al.}(2015){Geier}, {F{\"u}rst}, {Ziegerer}, {Kupfer},
  {Heber}, {Irrgang}, {Wang}, {Liu}, {Han}, {Sesar}, {Levitan}, {Kotak},
  {Magnier}, {Smith}, {Burgett}, {Chambers}, {Flewelling}, {Kaiser},
  {Wainscoat}, \& {Waters}}]{2015Sci...347.1126G}
{Geier}, S., {F{\"u}rst}, F., {Ziegerer}, E., {et~al.} 2015, Science, 347, 1126

\bibitem[{{Guillochon} {et~al.}(2010){Guillochon}, {Dan}, {Ramirez-Ruiz}, \&
  {Rosswog}}]{Guillochon}
{Guillochon}, J., {Dan}, M., {Ramirez-Ruiz}, E., \& {Rosswog}, S. 2010, \apjl,
  709, L64

\bibitem[{{Hills}(1988)}]{Hills1988}
{Hills}, J.~G. 1988, \nat, 331, 687

\bibitem[{{Hirsch} {et~al.}(2005){Hirsch}, {Heber}, {O'Toole}, \&
  {Bresolin}}]{2005A&A...444L..61H}
{Hirsch}, H.~A., {Heber}, U., {O'Toole}, S.~J., \& {Bresolin}, F. 2005, \aap,
  444, L61

\bibitem[{{Iglesias} \& {Rogers}(1993{\natexlab{a}})}]{Iglesias1993}
{Iglesias}, C.~A. \& {Rogers}, F.~J. 1993{\natexlab{a}}, \apj, 412, 752

\bibitem[{{Iglesias} \& {Rogers}(1993{\natexlab{b}})}]{1993ApJ...412..752I}
{Iglesias}, C.~A. \& {Rogers}, F.~J. 1993{\natexlab{b}}, \apj, 412, 752

\bibitem[{{Iglesias} \& {Rogers}(1996)}]{Iglesias1996}
{Iglesias}, C.~A. \& {Rogers}, F.~J. 1996, \apj, 464, 943

\bibitem[{{Irwin}(2004)}]{Irwin2004}
{Irwin}, A.~W. 2004, The FreeEOS Code for Calculating the Equation of State for
  Stellar Interiors

\bibitem[{{Itoh} {et~al.}(1996){Itoh}, {Hayashi}, {Nishikawa}, \&
  {Kohyama}}]{Itoh1996}
{Itoh}, N., {Hayashi}, H., {Nishikawa}, A., \& {Kohyama}, Y. 1996, \apjs, 102,
  411

\bibitem[{{Jermyn} {et~al.}(2023){Jermyn}, {Bauer}, {Schwab}, {Farmer}, {Ball},
  {Bellinger}, {Dotter}, {Joyce}, {Marchant}, {Mombarg}, {Wolf}, {Sunny Wong},
  {Cinquegrana}, {Farrell}, {Smolec}, {Thoul}, {Cantiello}, {Herwig}, {Toloza},
  {Bildsten}, {Townsend}, \& {Timmes}}]{Jermyn2023}
{Jermyn}, A.~S., {Bauer}, E.~B., {Schwab}, J., {et~al.} 2023, \apjs, 265, 15

\bibitem[{{Jermyn} {et~al.}(2021){Jermyn}, {Schwab}, {Bauer}, {Timmes}, \&
  {Potekhin}}]{Jermyn2021}
{Jermyn}, A.~S., {Schwab}, J., {Bauer}, E., {Timmes}, F.~X., \& {Potekhin},
  A.~Y. 2021, \apj, 913, 72

\bibitem[{{Kerzendorf} {et~al.}(2014){Kerzendorf}, {Childress},
  {Scharw{\"a}chter}, {Do}, \& {Schmidt}}]{2014ApJ...782...27K}
{Kerzendorf}, W.~E., {Childress}, M., {Scharw{\"a}chter}, J., {Do}, T., \&
  {Schmidt}, B.~P. 2014, \apj, 782, 27

\bibitem[{{Kerzendorf} {et~al.}(2018){Kerzendorf}, {Strampelli}, {Shen},
  {Schwab}, {Pakmor}, {Do}, {Buchner}, \& {Rest}}]{2018MNRAS.479..192K}
{Kerzendorf}, W.~E., {Strampelli}, G., {Shen}, K.~J., {et~al.} 2018, \mnras,
  479, 192

\bibitem[{{Kerzendorf} {et~al.}(2013){Kerzendorf}, {Yong}, {Schmidt}, {Simon},
  {Jeffery}, {Anderson}, {Podsiadlowski}, {Gal-Yam}, {Silverman}, {Filippenko},
  {Nomoto}, {Murphy}, {Bessell}, {Venn}, \& {Foley}}]{2013ApJ...774...99K}
{Kerzendorf}, W.~E., {Yong}, D., {Schmidt}, B.~P., {et~al.} 2013, \apj, 774, 99

\bibitem[{{Kippenhahn} {et~al.}(1980){Kippenhahn}, {Ruschenplatt}, \&
  {Thomas}}]{1980A&A....91..175K}
{Kippenhahn}, R., {Ruschenplatt}, G., \& {Thomas}, H.~C. 1980, \aap, 91, 175

\bibitem[{{Kromer} {et~al.}(2010){Kromer}, {Sim}, {Fink}, {R{\"o}pke},
  {Seitenzahl}, \& {Hillebrandt}}]{2010ApJ...719.1067K}
{Kromer}, M., {Sim}, S.~A., {Fink}, M., {et~al.} 2010, \apj, 719, 1067

\bibitem[{{Langanke} \& {Mart{\'{\i}}nez-Pinedo}(2000)}]{Langanke2000}
{Langanke}, K. \& {Mart{\'{\i}}nez-Pinedo}, G. 2000, Nuclear Physics A, 673,
  481

\bibitem[{{Lawlor} \& {MacDonald}(2006)}]{2006MNRAS.371..263L}
{Lawlor}, T.~M. \& {MacDonald}, J. 2006, \mnras, 371, 263

\bibitem[{{Li} \& {Zhao}(2017)}]{2017ApJ...850...25L}
{Li}, C. \& {Zhao}, G. 2017, \apj, 850, 25

\bibitem[{{Moll} \& {Woosley}(2013)}]{2013ApJ...774..137M}
{Moll}, R. \& {Woosley}, S.~E. 2013, \apj, 774, 137

\bibitem[{{Oda} {et~al.}(1994){Oda}, {Hino}, {Muto}, {Takahara}, \&
  {Sato}}]{Oda1994}
{Oda}, T., {Hino}, M., {Muto}, K., {Takahara}, M., \& {Sato}, K. 1994, Atomic
  Data and Nuclear Data Tables, 56, 231

\bibitem[{{Pakmor} {et~al.}(2022){Pakmor}, {Callan}, {Collins}, {de Mink},
  {Holas}, {Kerzendorf}, {Kromer}, {Neunteufel}, {O'Brien}, {R{\"o}pke},
  {Ruiter}, {Seitenzahl}, {Shingles}, {Sim}, \&
  {Taubenberger}}]{2022MNRAS.517.5260P}
{Pakmor}, R., {Callan}, F.~P., {Collins}, C.~E., {et~al.} 2022, \mnras, 517,
  5260

\bibitem[{{Pakmor} {et~al.}(2013){Pakmor}, {Kromer}, {Taubenberger}, \&
  {Springel}}]{2013ApJ...770L...8P}
{Pakmor}, R., {Kromer}, M., {Taubenberger}, S., \& {Springel}, V. 2013, \apjl,
  770, L8

\bibitem[{{Pakmor} {et~al.}(2016){Pakmor}, {Springel}, {Bauer}, {Mocz},
  {Munoz}, {Ohlmann}, {Schaal}, \& {Zhu}}]{Pakmor2016}
{Pakmor}, R., {Springel}, V., {Bauer}, A., {et~al.} 2016, \mnras, 455, 1134

\bibitem[{{Paxton} {et~al.}(2011){Paxton}, {Bildsten}, {Dotter}, {Herwig},
  {Lesaffre}, \& {Timmes}}]{Paxton2011}
{Paxton}, B., {Bildsten}, L., {Dotter}, A., {et~al.} 2011, \apjs, 192, 3

\bibitem[{{Paxton} {et~al.}(2013){Paxton}, {Cantiello}, {Arras}, {Bildsten},
  {Brown}, {Dotter}, {Mankovich}, {Montgomery}, {Stello}, {Timmes}, \&
  {Townsend}}]{Paxton2013}
{Paxton}, B., {Cantiello}, M., {Arras}, P., {et~al.} 2013, \apjs, 208, 4

\bibitem[{{Paxton} {et~al.}(2015){Paxton}, {Marchant}, {Schwab}, {Bauer},
  {Bildsten}, {Cantiello}, {Dessart}, {Farmer}, {Hu}, {Langer}, {Townsend},
  {Townsley}, \& {Timmes}}]{Paxton2015}
{Paxton}, B., {Marchant}, P., {Schwab}, J., {et~al.} 2015, \apjs, 220, 15

\bibitem[{{Paxton} {et~al.}(2018){Paxton}, {Schwab}, {Bauer}, {Bildsten},
  {Blinnikov}, {Duffell}, {Farmer}, {Goldberg}, {Marchant}, {Sorokina},
  {Thoul}, {Townsend}, \& {Timmes}}]{Paxton2018}
{Paxton}, B., {Schwab}, J., {Bauer}, E.~B., {et~al.} 2018, \apjs, 234, 34

\bibitem[{{Paxton} {et~al.}(2019){Paxton}, {Smolec}, {Schwab}, {Gautschy},
  {Bildsten}, {Cantiello}, {Dotter}, {Farmer}, {Goldberg}, {Jermyn}, {Kanbur},
  {Marchant}, {Thoul}, {Townsend}, {Wolf}, {Zhang}, \& {Timmes}}]{Paxton2019}
{Paxton}, B., {Smolec}, R., {Schwab}, J., {et~al.} 2019, \apjs, 243, 10

\bibitem[{{Poutanen}(2017)}]{Poutanen2017}
{Poutanen}, J. 2017, \apj, 835, 119

\bibitem[{{Raddi} {et~al.}(2018){Raddi}, {Hollands}, {Koester}, {G{\"a}nsicke},
  {Gentile Fusillo}, {Hermes}, \& {Townsley}}]{Raddi2018}
{Raddi}, R., {Hollands}, M.~A., {Koester}, D., {et~al.} 2018, \apj, 858, 3

\bibitem[{{Raddi} {et~al.}(2019){Raddi}, {Hollands}, {Koester}, {Hermes},
  {G{\"a}nsicke}, {Heber}, {Shen}, {Townsley}, {Pala}, {Reding}, {Toloza},
  {Pelisoli}, {Geier}, {Gentile Fusillo}, {Munari}, \&
  {Strader}}]{2019MNRAS.489.1489R}
{Raddi}, R., {Hollands}, M.~A., {Koester}, D., {et~al.} 2019, \mnras, 489, 1489

\bibitem[{{Schwab} {et~al.}(2012){Schwab}, {Shen}, {Quataert}, {Dan}, \&
  {Rosswog}}]{Schwab2012}
{Schwab}, J., {Shen}, K.~J., {Quataert}, E., {Dan}, M., \& {Rosswog}, S. 2012,
  \mnras, 427, 190

\bibitem[{{Shen} {et~al.}(2012){Shen}, {Bildsten}, {Kasen}, \&
  {Quataert}}]{Shen2012}
{Shen}, K.~J., {Bildsten}, L., {Kasen}, D., \& {Quataert}, E. 2012, \apj, 748,
  35

\bibitem[{{Shen} {et~al.}(2024){Shen}, {Boos}, \& {Townsley}}]{Shen2024}
{Shen}, K.~J., {Boos}, S.~J., \& {Townsley}, D.~M. 2024, \apj, 975, 127

\bibitem[{{Shen} {et~al.}(2018{\natexlab{a}}){Shen}, {Boubert}, {G{\"a}nsicke},
  {Jha}, {Andrews}, {Chomiuk}, {Foley}, {Fraser}, {Gromadzki}, {Guillochon},
  {Kotze}, {Maguire}, {Siebert}, {Smith}, {Strader}, {Badenes}, {Kerzendorf},
  {Koester}, {Kromer}, {Miles}, {Pakmor}, {Schwab}, {Toloza}, {Toonen},
  {Townsley}, \& {Williams}}]{2018ApJ...865...15S}
{Shen}, K.~J., {Boubert}, D., {G{\"a}nsicke}, B.~T., {et~al.}
  2018{\natexlab{a}}, \apj, 865, 15

\bibitem[{{Shen} {et~al.}(2018{\natexlab{b}}){Shen}, {Kasen}, {Miles}, \&
  {Townsley}}]{Shen2018a}
{Shen}, K.~J., {Kasen}, D., {Miles}, B.~J., \& {Townsley}, D.~M.
  2018{\natexlab{b}}, \apj, 854, 52

\bibitem[{{Shen} \& {Schwab}(2017)}]{Shen2017}
{Shen}, K.~J. \& {Schwab}, J. 2017, \apj, 834, 180

\bibitem[{Shields {et~al.}(2023)Shields, Arunachalam, Kerzendorf, Hughes,
  Biriouk, Monk, \& Buchner}]{2023ApJ...950L..10S}
Shields, J.~V., Arunachalam, P., Kerzendorf, W., {et~al.} 2023, \apj, 950, L10

\bibitem[{Shields {et~al.}(2022)Shields, Kerzendorf, Hosek, Shen, Rest, Do, Lu,
  Fullard, Strampelli, \& Zenteno}]{2022ApJ...933L..31S}
Shields, J.~V., Kerzendorf, W., Hosek, M.~W., {et~al.} 2022, \apj, 933, L31

\bibitem[{{Springel}(2010)}]{Arepo}
{Springel}, V. 2010, \mnras, 401, 791

\bibitem[{{Tanikawa} {et~al.}(2019){Tanikawa}, {Nomoto}, {Nakasato}, \&
  {Maeda}}]{2019ApJ...885..103T}
{Tanikawa}, A., {Nomoto}, K., {Nakasato}, N., \& {Maeda}, K. 2019, \apj, 885,
  103

\bibitem[{{Timmes} \& {Swesty}(2000{\natexlab{a}})}]{Timmes2000}
{Timmes}, F.~X. \& {Swesty}, F.~D. 2000{\natexlab{a}}, \apjs, 126, 501

\bibitem[{{Timmes} \& {Swesty}(2000{\natexlab{b}})}]{2000ApJS..126..501T}
{Timmes}, F.~X. \& {Swesty}, F.~D. 2000{\natexlab{b}}, \apjs, 126, 501

\bibitem[{{Vennes} {et~al.}(2017){Vennes}, {Nemeth}, {Kawka}, {Thorstensen},
  {Khalack}, {Ferrario}, \& {Alper}}]{Vennes2017}
{Vennes}, S., {Nemeth}, P., {Kawka}, A., {et~al.} 2017, Science, 357, 680

\bibitem[{{Weinberger} {et~al.}(2020){Weinberger}, {Springel}, \&
  {Pakmor}}]{2020ApJS..248...32W}
{Weinberger}, R., {Springel}, V., \& {Pakmor}, R. 2020, \apjs, 248, 32

\bibitem[{{Werner} {et~al.}(2024){Werner}, {Reindl}, {Rauch}, {El-Badry}, \&
  {B{\'e}dard}}]{2024A&A...682A..42W}
{Werner}, K., {Reindl}, N., {Rauch}, T., {El-Badry}, K., \& {B{\'e}dard}, A.
  2024, \aap, 682, A42

\bibitem[{{Wong} {et~al.}(2024){Wong}, {White}, \& {Bildsten}}]{sunny2024}
{Wong}, T. L.~S., {White}, C.~J., \& {Bildsten}, L. 2024, \apj, 973, 65

\bibitem[{{Zhang} {et~al.}(2019){Zhang}, {Fuller}, {Schwab}, \&
  {Foley}}]{2019ApJ...872...29Z}
{Zhang}, M., {Fuller}, J., {Schwab}, J., \& {Foley}, R.~J. 2019, \apj, 872, 29

\end{thebibliography}
\clearpage
\appendix

\section{Comparing the relaxed and heated Arepo remnant}
\label{A.Comparison}
\begin{figure}[h!]
\centering
\includegraphics[width=0.455\textwidth]{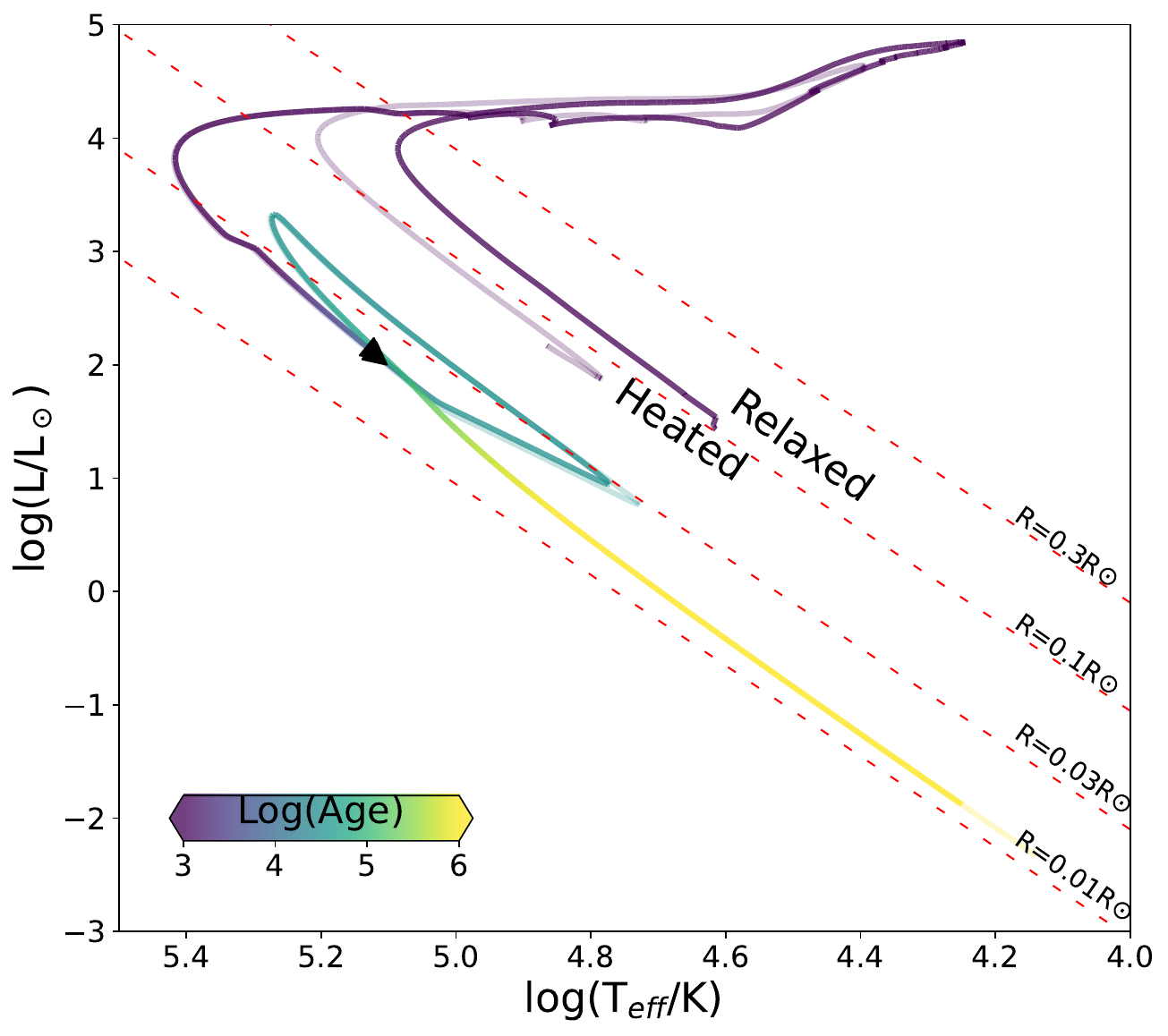} 
\includegraphics[width=0.45\textwidth]
{plots_second_version/default_evolution_tailless.pdf}
\caption{\textbf{[Top Panel]}Evolutionary track of the $0.64 M_\odot$ model relaxed directly to the remnant profile and the model relaxed to donor profile and then heated. Red lines show lines of constant radii from $0.01-0.3 R_\odot$. Only $\sim 0.001\%$ of the total evolutionary time is spent above $R>0.03 R_\odot$, without considering the helium flash. \textbf{[Bottom Panel]} A comparison plot of the model without any helium in the inner layers of the star.}
\label{fig:def_evolve}
\end{figure}
We show evolutionary tracks in the HR diagram for the $0.64\,M_\odot$ model mapped directly from Arepo in the left panel of Figure~\ref{fig:def_evolve}. The bump in luminosity and radius at around $3\times10^4$ years occurs due to unstable burning of helium in both relaxed and heated cases. As we do not think the presence of helium is physically realistic, we also show a comparison with the model without any helium in the inner regions in the right panel of Figure~\ref{fig:def_evolve}.

\begin{figure}
\centering
\hspace*{-0.2cm}
\includegraphics[width=0.45\textwidth]{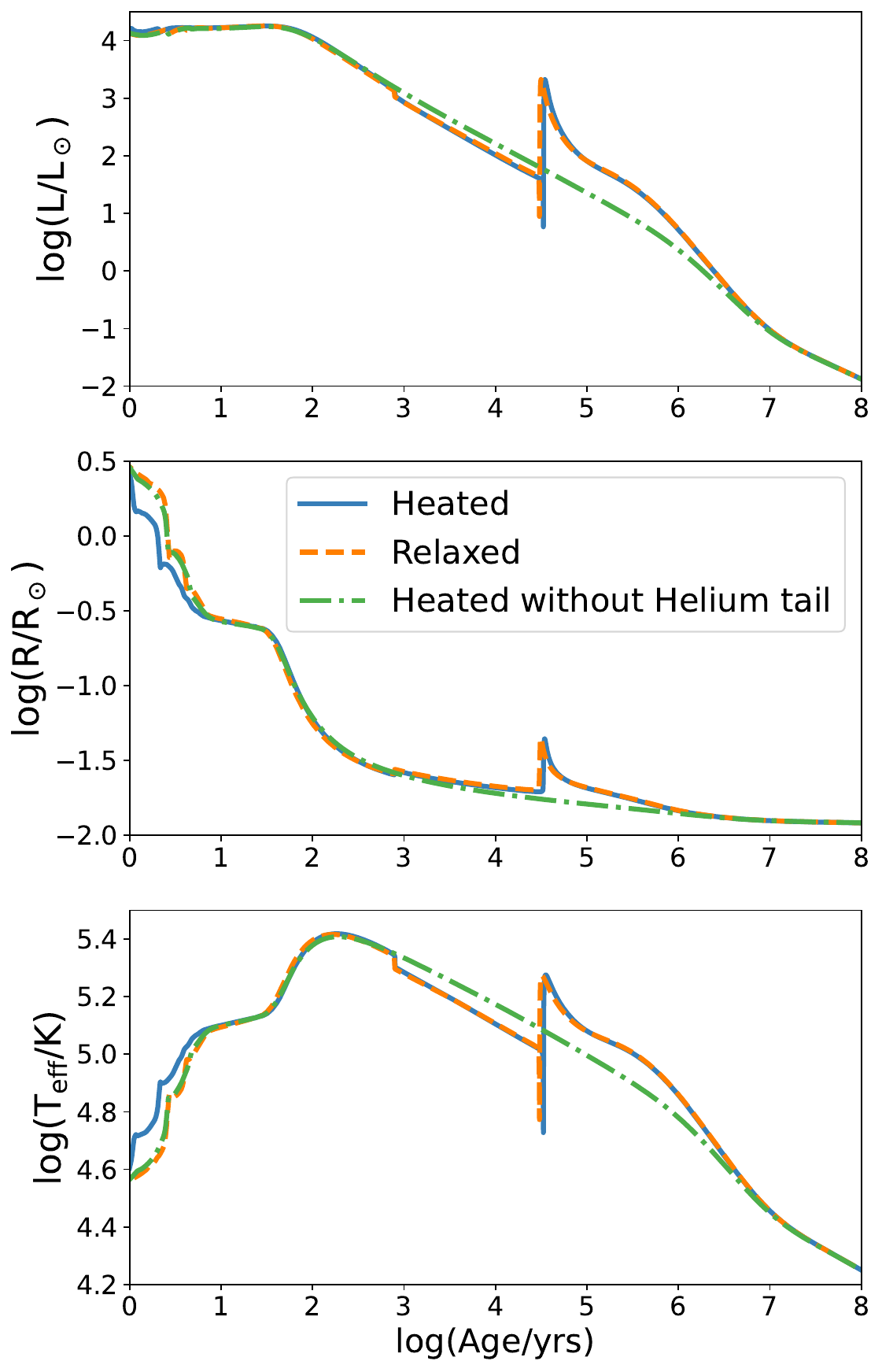} 
\caption{Evolution of surface luminosity, radius, and effective temperature of the heated (blue) and relaxed (orange) models for comparison. Also plotted are the quantities for the heated model without any helium in the inner regions of the star (green).}
\label{fig:def_evolve_lrt}
\end{figure}

 Both the relaxed and heated remnant models reach a similar state quickly. We show the evolution of stellar surface quantities in Figure~\ref{fig:def_evolve_lrt}. There are no significant differences between the two models. For the model without helium inside, the track converges to the same state as the other models after a few Myr. The models reach super-Eddington luminosity due to heating and Nickel decay within a few days. The Eddington limit is reached within the first year but declines quickly after the first 100 years. The star briefly inflates to a large radius, but this state is short-lived. The bumps in all three quantities after $10^4 \mathrm{yrs}$ for the first two models are due to a helium flash. Those models ignite this helium at the same time, confirming that the initial states are the same for the relaxation and heating. This is not seen for the model without helium in the inner 99\% of the star. 

 This comparison shows that heating the model provides the same result as a direct relaxation, in the case when the compositions are the same, and the heat is calculated as a difference of entropy from the \textsc{Arepo} model. We therefore are able to generalise the heating procedure and rely on it for the evolution of the bigger grid of models. Insofar as the temperature profiles of the white dwarfs are roughly the same, this method can be applied to a model of any mass. Furthermore the end state of the relaxed model is the same as the end state of the helium-less heated model, except for a short-lived loop in the HR diagram. After a few thousand years both of these models are in the same region of the HR diagram.

\section{Models with \textsc{Arepo} composition}
\label{B.Arepo}

\begin{figure*}
\centering
\includegraphics[width=0.46\textwidth]{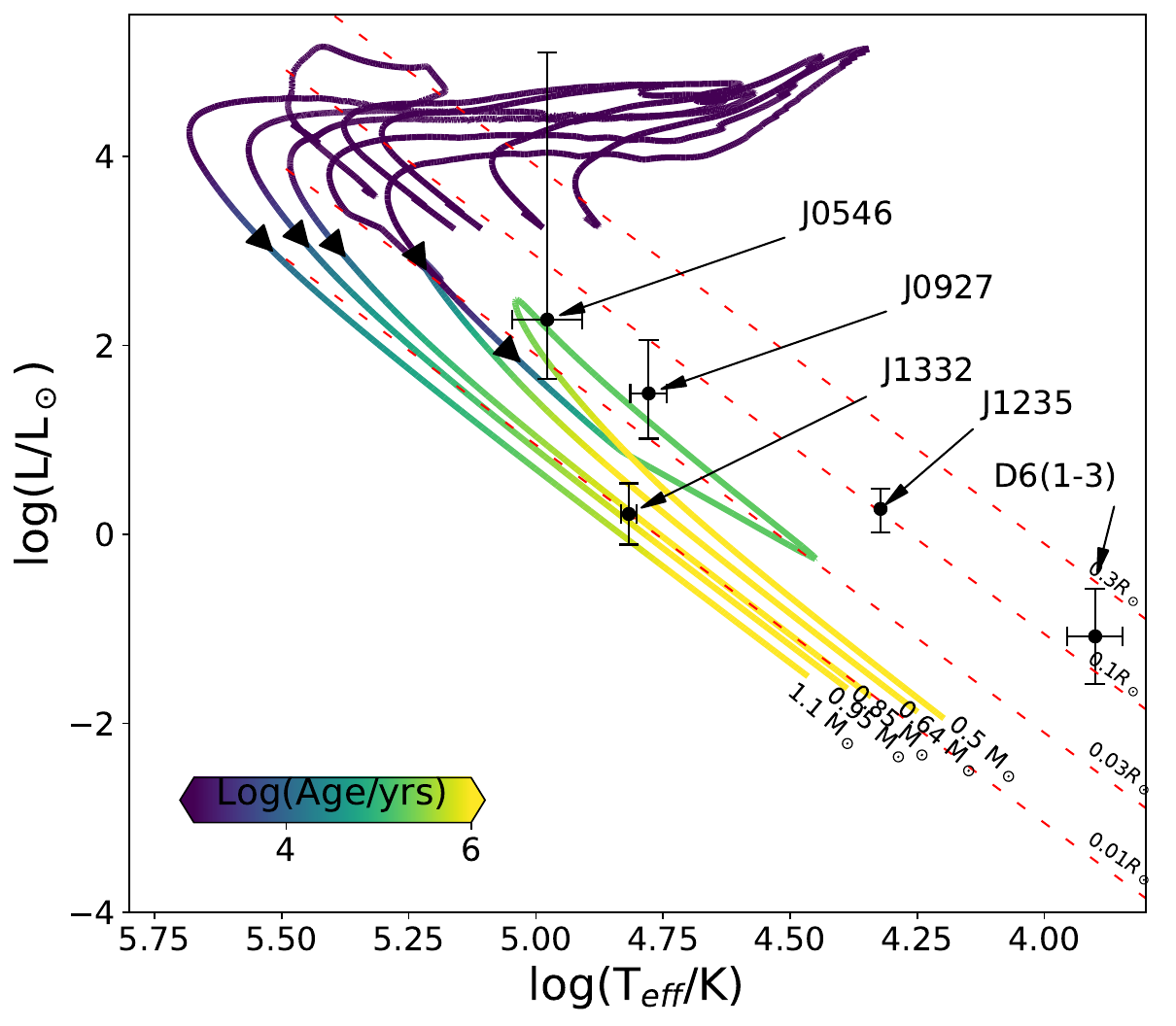} 
\includegraphics[width=0.45\textwidth]{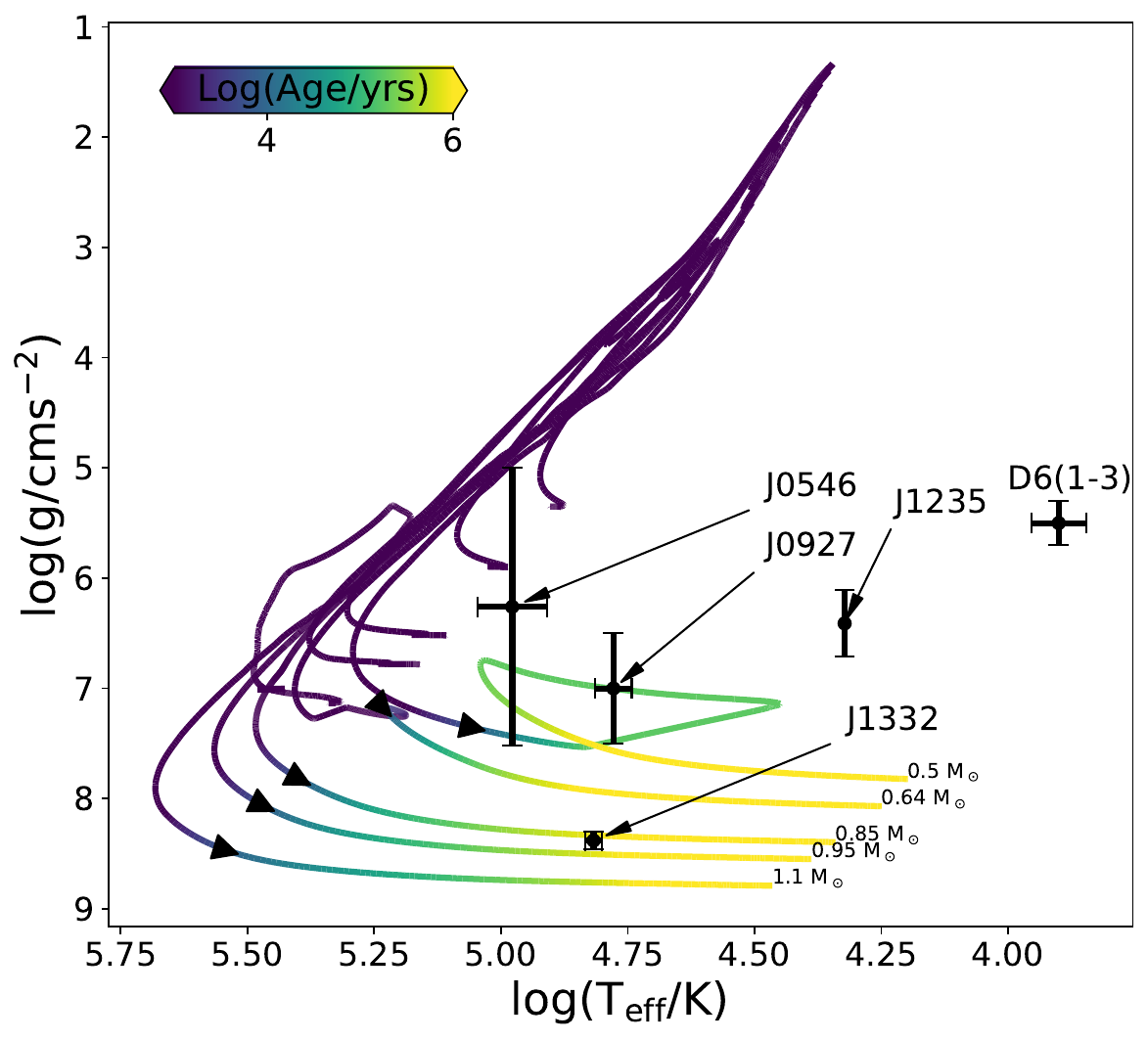}
\caption{\textbf{[left panel]} HR diagram for grid 1 models. All reported D$^6$ stars are plotted as black points. Red lines represent lines of constant radii. Colorbar shows the log(age) of the star. \textbf{[right panel]} The Kiel diagram for the same model. The high uncertainty in J3546 comes from combining the minimum and maximum values from \citet{2024A&A...682A..42W} and \citet{2023arXiv230603914E}. The big loop of $0.50M_\odot$ and the smaller loop of $0.64M_\odot$  models is due to unstable helium burning.}
\label{fig:evolve_helium}
\end{figure*}

 The HR diagram of all the models with Arepo composition (no re-scaling of helium) is shown in Figure \ref{fig:evolve_helium}, with the D6 observations over-plotted. These models have a total of around 0.002 M$_\odot$ of Helium.

The HR diagram of all models is similar to the models without helium discussed before. All stars spend $<10^4 \mathrm{yrs}$ at high luminosities, by which time the Nickel decay is quenched and the star has lost some mass due to radiation powered winds. After this time the stars continue to burn helium stably, and reach the tip of the white dwarf cooling track. There is only one model where the helium flash occurs. However, since this only occurs in the lowest mass WD, it should not be considered as an explanation for the observed stars. This is especially because the D$^6$ stars lying on this region of the HR diagram should have been high mass ($>0.8 M_\odot$) donors \citep{2021ApJ...923L..34B}. This instability is also dependent on the initial structure of the chosen model, as the 0.64 $M_\odot$ model does not have a strong helium flash in this case (as opposed to the relaxed models before). 
For a better comparison with spectral results, we also show the Kiel diagram in right panel of Figure \ref{fig:evolve_helium}. 

\begin{figure}
\centering
\hspace*{-0.2cm}
\includegraphics[width=0.45\textwidth]{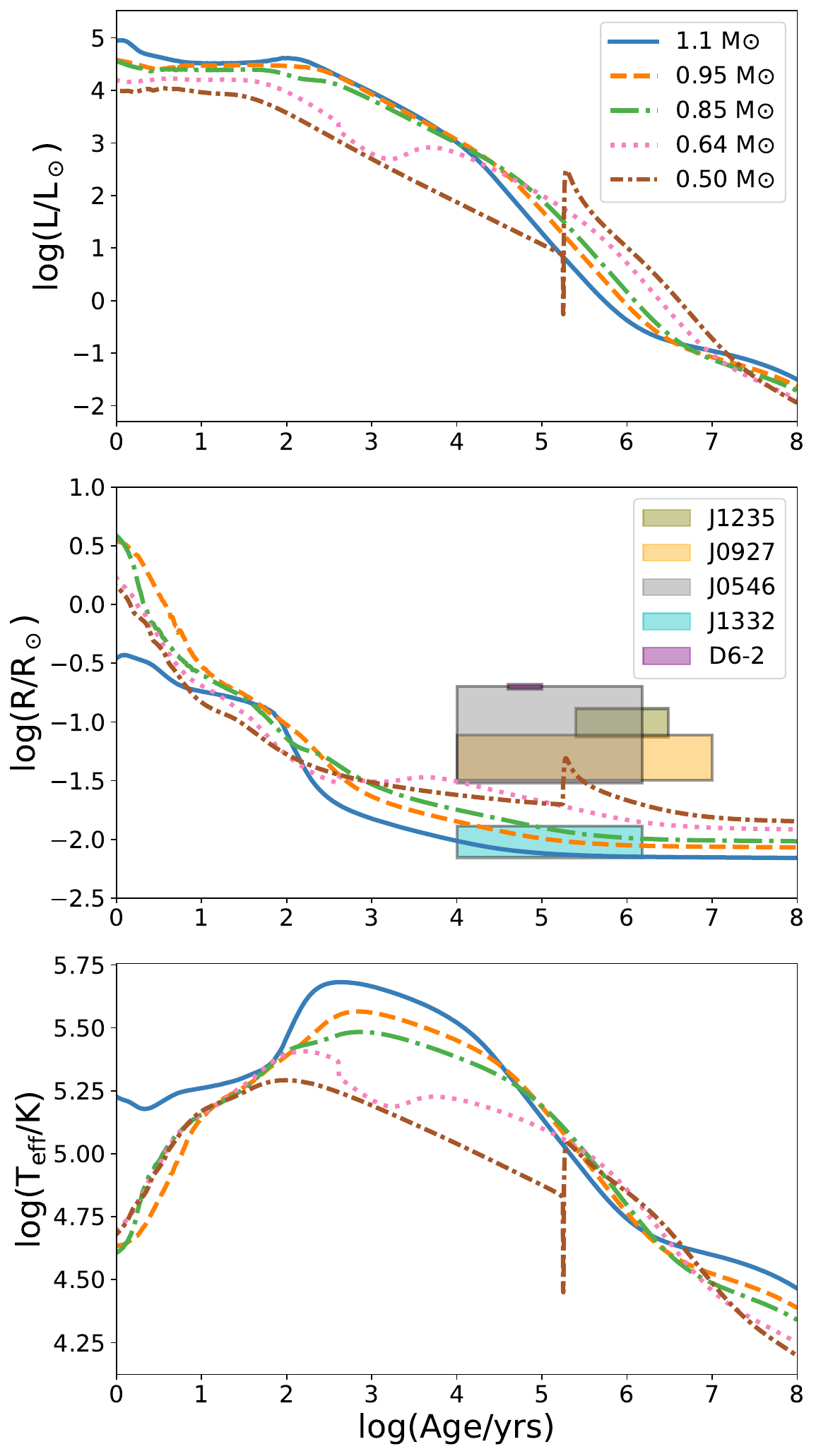} 
\caption{Evolution of the luminosity, radius, and temperature of models with \textsc{Arepo} helium content}
\label{fig:lrt_1}
\end{figure}

The evolution of surface luminosity, radius, and temperature are shown in Figure \ref{fig:lrt_1}. We also plot regions of estimated radii of known D$^6$ stars for an age between $10^4-10^7 \mathrm{yrs}$. The fact that two D$^6$ stars coincide with the radii of the 0.5 $\mathrm{M}_\odot$ model is again due to the helium flash and should not be considered as a confirmation of the models. 







%
%



\end{document}